\documentclass[journal,comsoc]{IEEEtran}
\usepackage[T1]{fontenc}

\usepackage{tabularx}
\usepackage{xcolor}
\usepackage{multirow}
\usepackage{booktabs}
\usepackage{array}
\newcolumntype{L}[1]{>{\raggedright\let\newline\\\arraybackslash\hspace{0pt}}m{#1}}
\newcolumntype{C}[1]{>{\centering\let\newline\\\arraybackslash\hspace{0pt}}m{#1}}
\newcolumntype{R}[1]{>{\raggedleft\let\newline\\\arraybackslash\hspace{0pt}}m{#1}}

\usepackage[english]{babel}
\usepackage{amsthm}
\usepackage{amssymb}

\newtheorem{theorem}{Theorem}
\newtheorem{lemma}{Lemma}

\theoremstyle{plain}
\newtheorem{proposition}{Proposition}
\theoremstyle{remark}

\usepackage{cite}

\ifCLASSINFOpdf
   \usepackage[pdftex]{graphicx}
\else
   or other class option (dvipsone, dvipdf, if not using dvips). graphicx
   \usepackage[dvips]{graphicx}
   \graphicspath{{../eps/}}
   \DeclareGraphicsExtensions{.eps}
\fi
\usepackage[normalem]{ulem}

\usepackage{amsmath}
\usepackage[cmintegrals]{newtxmath}
\usepackage{algorithmic}
\usepackage{algorithm}

\usepackage{verbatim}

\usepackage{array}
\usepackage{makecell}

\usepackage[pagewise]{lineno}
\usepackage{upgreek}

\ifCLASSOPTIONcompsoc
  \usepackage[caption=false,font=normalsize,uabelfont=sf,textfont=sf]{subfig}
\else
  \usepackage[caption=false,font=footnotesize]{subfig}
\fi
\begin{document}
\abovedisplayshortskip=1pt
\belowdisplayshortskip=1pt
\abovedisplayskip=1pt
\belowdisplayskip=1pt
\textfloatsep=1pt
\floatsep=1pt
\intextsep=1pt

\setcounter{figure}{0}
\renewcommand{\figurename}{Fig.}
\renewcommand{\thefigure}{\arabic{figure}}
\title{Over-the-Air Time-Frequency Synchronization in Distributed ISAC Systems}
%
%

\author{Kawon~Han,~\IEEEmembership{Member,~IEEE,} Kaitao~Meng,~\IEEEmembership{Member,~IEEE,} and Christos~Masouros,~\IEEEmembership{Fellow,~IEEE}

\thanks{Manuscript received xx, 2025. K. Han, K. Meng, and C. Masouros are with the Department of Electronic and Electrical Engineering, University College London, London, UK (emails: {kawon.han, kaitao.meng, c.masouros}@ucl.ac.uk).}}

\maketitle

\begin{abstract}
A distributed integrated sensing and communication (D-ISAC) system offers significant cooperative gains for both sensing and communication performance. These gains, however, can only be fully realized when the distributed nodes are perfectly synchronized, which is a challenge that remains largely unaddressed in current ISAC research. In this paper, we propose an over-the-air time-frequency synchronization framework for the D-ISAC system, leveraging the reciprocity of bistatic sensing channels. This approach overcomes the impractical dependency of traditional methods on a direct line-of-sight (LoS) link, enabling the estimation of time offset (TO) and carrier frequency offset (CFO) between two ISAC nodes even in non-LoS (NLOS) scenarios. To achieve this, we introduce a bistatic signal matching (BSM) technique with delay-Doppler decoupling, which exploits offset reciprocity (OR) in bistatic observations. This method compresses multiple sensing links into a single offset for estimation. We further present off-grid super-resolution estimators for TO and CFO, including the maximum likelihood estimator (MLE) and the matrix pencil (MP) method, combined with BSM processing. These estimators provide more accurate offset estimation compared to spectral cross-correlation techniques based on on-grid estimators. Additionally, we extend the pairwise synchronization leveraging OR between two nodes to the synchronization of $N$ multiple distributed nodes, referred to as centralized pairwise synchronization. We analyze the Cramér-Rao bounds (CRBs) for TO and CFO estimates and evaluate the impact of D-ISAC synchronization on the bottom-line target localization performance. Simulation results validate the effectiveness of the proposed algorithm, confirm the theoretical analysis, and demonstrate that the proposed synchronization approach can recover up to 96$\%$ of the bottom-line target localization performance of the fully-synchronous D-ISAC. 
\end{abstract}


%
\IEEEpeerreviewmaketitle

\section{Introduction}
%
%
%
%
\IEEEPARstart{I}ntegrated sensing and communication (ISAC) has emerged as a key technology for next-generation wireless networks, enabling the joint delivery of sensing and communication services through shared hardware and spectrum \cite{liu2022integrated, ISAC1, ISAC3}. By optimizing infrastructure, frequency spectrum, and power allocation of ISAC systems, it fosters sustainable development of future wireless networks \cite{zhu2024enabling, valiulahi2023net, zhang2021enabling}. This has attracted significant attention, driving advancements in signaling design \cite{liu2020joint}, waveform selection \cite{rou2024otfs}, resource allocation \cite{dong2022sensing}, and physical layer security \cite{su2020secure}.

The ultimate goal of deploying ISAC in cellular networks is to enable coordinated sensing of unprecedented scale. The growing density of networks introduces challenges such as inter-cell interference and resource congestion, which limit the effectiveness of multi-cell systems that rely on schemes originally designed for single-node systems \cite{salem2024rethinking}. To address these issues, networked ISAC systems, which integrate sensing and communication functionalities cooperatively, have emerged as a promising solution \cite{meng2024cooperative, meng2024network}. These systems range from multi-cell cooperation \cite{babu2024precoding} to cell-free multi-input multi-output (MIMO) configurations \cite{demirhan2024cell}, leveraging distributed basestation (BS) or access point (AP) nodes. In communication, multi-node cooperation can mitigate mutual interference via coordinated beamforming \cite{dahrouj2010coordinated} and achieves signal combining gains through joint transmission coordinated multipoint (CoMP) \cite{irmer2011coordinated}. On the sensing side, distributed MIMO radar improves target localization by exploiting enhanced spatial diversity \cite{godrich2010target}. Consequently, distributed ISAC (D-ISAC) significantly enhances both sensing and communication performance, surpassing the capabilities of single-node ISAC systems.

Recent studies have examined the performance gains of D-ISAC in various scenarios, including signaling design in D-ISAC \cite{han2025signaling}, beamforming for multi-static sensing and CoMP transmission \cite{xu2023integrated}, and target localization using cell-free cooperative ISAC \cite{zhang2024target}. Crucially, without robust synchronization, distributed nodes suffer from misaligned timing, frequency, and phase errors that dramatically erode both sensing accuracy and communication efficiency \cite{strinati2024towards}. However, synchronization among cooperative ISAC nodes has received limited attention, with few studies addressing its impact on ISAC performance. As a pioneering study, the impact of time synchronization errors on target detection performance in networked ISAC systems was investigated using coordinated transmit beamforming \cite{cheng2024optimal}. While this study underscores the importance of synchronization in enhancing target detection probability for D-ISAC, it focuses solely on monostatic links in unsynchronized scenarios, neglecting performance degradation in bistatic measurements. Similarly, \cite{yang2024coordinated} proposed a coordinated transmit beamforming approach based on known synchronization error statistics in systems with separately deployed transmitting and receiving BSs. However, solutions for achieving synchronization in D-ISAC networks remain unexplored.

\subsection{State-of-the-Art of Over-the-Air Synchronization}
As synchronization in D-ISAC is critical for fully realizing cooperation gains in both wireless communications and radar sensing, this section investigates existing over-the-air synchronization techniques and highlights the unique opportunities that ISAC systems offer for D-ISAC synchronization. Mitigating time offsets (TOs) and carrier frequency offsets (CFOs) to synchronize physically separated nodes using over-the-air transmission is inherently challenging due to the availability of line-of-sight (LoS) links and limited signal-to-noise ratio (SNR). For instance, global navigation satellite systems (GNSS), such as the global positioning system (GPS), can provide time-frequency synchronization for distributed wireless networks with an accuracy on the order of 100 ns \cite{zou2015network}. However, GNSS-based synchronization exhibits poor performance in indoor and dense urban environments \cite{lasassmeh2010time}.

To address the limitations of satellite-based synchronization, self-synchronization solutions for distributed wireless networks have been extensively studied \cite{mudumbai2007feasibility, cox2005time, elson2002fine, ganeriwal2003timing}. A widely used approach is the master-slave configuration, where a master node broadcasts a training signal to synchronize slave nodes \cite{mudumbai2007feasibility, cox2005time}. Reference broadcast synchronization \cite{elson2002fine} compares receiver-to-receiver offsets using reference broadcasting but ignores the time-of-flight (TOF) of signals. Similarly, the Timing-Sync Protocol for Sensor Networks (TPSN) \cite{ganeriwal2003timing} achieves synchronization by exchanging timestamps over hierarchical distributed nodes through level discovery and synchronization phases. While these timestamp-based protocols achieve microsecond-level synchronization, they cannot achieve the cm-level localization accuracy required for high-precision radar sensing applications. Additionally, they rely on the assumption of strong LoS links between distributed nodes for broadcasting reference signals, which may not always be available.

Instead of transmitting unmodulated beacons, some synchronization methods utilize frequency-modulated continuous-wave (FMCW) \cite{roehr2007method}, two-tone waveforms \cite{merlo2022wireless}, and ultrawideband (UWB) pulses \cite{prager2020wireless}, leveraging their wideband characteristics for more accurate estimation of TO between nodes. Using two-way time transfer (TWTT) synchronization, distributed UWB radar networks have achieved subnanosecond time synchronization \cite{prager2020wireless}. Also, a cross-correlation based on the preamble and pilot signals using the strong LoS link has been utilized for over-the-air synchronization in bistatic ISAC \cite{brunner2024bistatic}. However, these methods also depend on the availability of LoS channels, which may not be consistently achievable in various ISAC application scenarios, such as indoor environments and vehicular networks. 

Recent studies have investigated non-line-of-sight (NLoS) synchronization techniques \cite{alemdar2021rfclock, merlo2023picosecond, pegoraro2024jump, wang2024clutter, jiang2024cooperation}. In \cite{alemdar2021rfclock}, a time-frequency reference source, RFClock, is used to synchronize distributed wireless nodes, requiring additional hardware dedicated solely to synchronization. The work in \cite{merlo2023picosecond} employs a known strong reflector to enable TWTT synchronization over NLoS links. Wireless synchronization methods for bistatic sensing without strong LoS links have also been explored. In \cite{pegoraro2024jump}, channel impulse response (CIR) and its cross-correlation across consecutive frames are used to estimate TOs between transmitters and receivers. Similarly, \cite{wang2024clutter} introduces a fingerprint-spectrum-based TO and CFO compensation technique, leveraging static clutter reflections as a reference spectrum. However, these methods primarily address TO and CFO drift between frames, failing to resolve initial TO and CFO differences critical for accurate target localization and velocity estimation. 

On the other hand, in terms of initial TO and CFO mitigation, the cross-correlation cooperative sensing (CCCS) technique proposed in \cite{jiang2024cooperation} treats the active monostatic sensing link as a reference to mitigate TO and CFO in passive bistatic sensing links. While this approach improves the ranging performance of the active sensing node by constructively exploiting passive link, it does not utilize the additional spatial diversity that can be harnessed from cooperative transmitting nodes for improved target localization. Furthermore, cooperative radar networks \cite{kong2013wireless, aguilar2024uncoupled, sigg2024over} estimate TO between uncoupled radars by comparing range measurements from two bistatic links. This method depends on the estimation of a single dominant target and does not account for multiple target reflections, limiting its effectiveness in complex scenarios containing multiple scatterers.

From the perspective of TO and CFO estimation, cross-correlation (CC) of delay-Doppler spectra has been widely used in synchronization approaches \cite{pegoraro2024jump, wang2024clutter, jiang2024cooperation, kong2013wireless, aguilar2024uncoupled, sigg2024over}. However, these methods have limited accuracy due to their on-grid property. Increasing the number of estimate grid points using zero-padding to improve accuracy significantly enlarges the computational complexity of the 2D CC spectrum. Although high-resolution spectral analysis methods like MUSIC can improve offset estimation, they also suffer from grid-constrained accuracy as they rely on spectrum-based approaches. An efficient off-grid super-resolution estimator is therefore needed to achieve tighter synchronization, which is crucial for maximizing the cooperative performance gains in D-ISAC.

\subsection{Our Contributions}
As fine-grained and robust synchronization is essential for both cooperative communication and sensing performance in D-ISAC, a high-accuracy TO and CFO estimation and compensation technique without reliance on LoS links is required. Interestingly, ISAC offers a novel opportunity for over-the-air synchronization in distributed nodes by leveraging sensing channels from multiple targets or scatterers for offset estimation, which is not available on the classical communication-only networks. 

Building on this insight, we propose an over-the-air time-frequency synchronization framework for D-ISAC that uses multiple sensing targets or scatterers as media for synchronization between distinct nodes. Our approach exploits the principle of reciprocal bistatic observations, wherein the TO and CFO between a node pair manifest as equal values but opposite directions, a phenomenon we define as offset reciprocity (OR). To enable super-resolution offset estimation (SOE) using OR, we develop a bistatic signal matching procedure that compresses the multiple sensing target links into a single offset component. This pairwise approach is then extended to cases involving multiple distributed nodes, with a tractable expression derived for the lower bound of the total offset estimation variance. The main contributions of this work are summarized as follows:
\begin{itemize}
    \item 
    We propose a novel over-the-air time-frequency synchronization technique that exploits multiple sensing targets or scatterers as synchronization media. Unlike traditional methods relying on LoS links or dedicated synchronization hardware, this technique leverages sensing information available in D-ISAC systems without requiring additional signal design, enabling robust synchronization in challenging environments.
    
    \item 
    We develop a super-resolution TO and CFO estimator based on OR, where bistatic observations between pairs of nodes are used to achieve fine-grained synchronization. This is developed with a bistatic signal matching (BSM) processing, which enables accurate offset estimation by compressing sensing target information into a single offset component. The maximum likelihood estimation (MLE) and matrix pencil (MP) method are leveraged as off-grid offset estimators, which overcomes the limitations of on-grid estimation methods using the spectral cross-correlation.
    
    \item 
    We extend the pairwise synchronization approach to an $N$-node distributed network, proposing a scalable framework for multi-node synchronization. Given randomly deployed ISAC nodes, we derive a tractable theoretical performance, providing a lower bound on the total offset estimation variance, which ensures a comprehensive evaluation of synchronization accuracy with the proposed approach.
    
    \item 
    We reveal the fundamental relationship between synchronization errors and the bottom-line performance of D-ISAC systems. By quantifying the impact of synchronization inaccuracies on cooperative sensing in D-ISAC, we establish the synchronization requirements necessary to unlock the full potential of D-ISAC networks.
\end{itemize}

$Notations$: Boldface variables with lower- and upper-case symbols represent vectors and matrices, respectively. $\textbf{A} \in \mathbb{C}^{N \times M}$ and $\textbf{B} \in \mathbb{R}^{N \times M}$ denotes a complex-valued ${N \times M}$ matrix $\textbf{A}$ and a real-valued ${N \times M}$ matrix $\textbf{B}$, respectively. Also, $\mathbf{0}_{N \times M}$ and $\mathbf{I}_N$ denote a $N \times M$ zero-matrix and a $N \times N$ identity matrix, respectively. $({\cdot})^{T}$, $({\cdot})^{H}$, and $({\cdot})^{*}$ represent the transpose, Hermitian transpose, and conjugate operators, respectively. ${\text{diag}({\textbf{a}})}$ denotes a diagonal matrix with diagonal entries of a vector $\textbf{a}$. The operator $\odot$ represents the Hadamard (element-wise) product. The function $\Pi(\cdot)$ denotes the rectangular pulse function. $\Re(a)$ and $\Im(a)$ represent the real and imaginary part of the complex number $a$, respectively. $\mathbb{E}{[\cdot]}$ is the statistical expectation operator. 

\section{System Model}
\begin{figure}[t!]
    \centering
    {\includegraphics[scale=0.40]{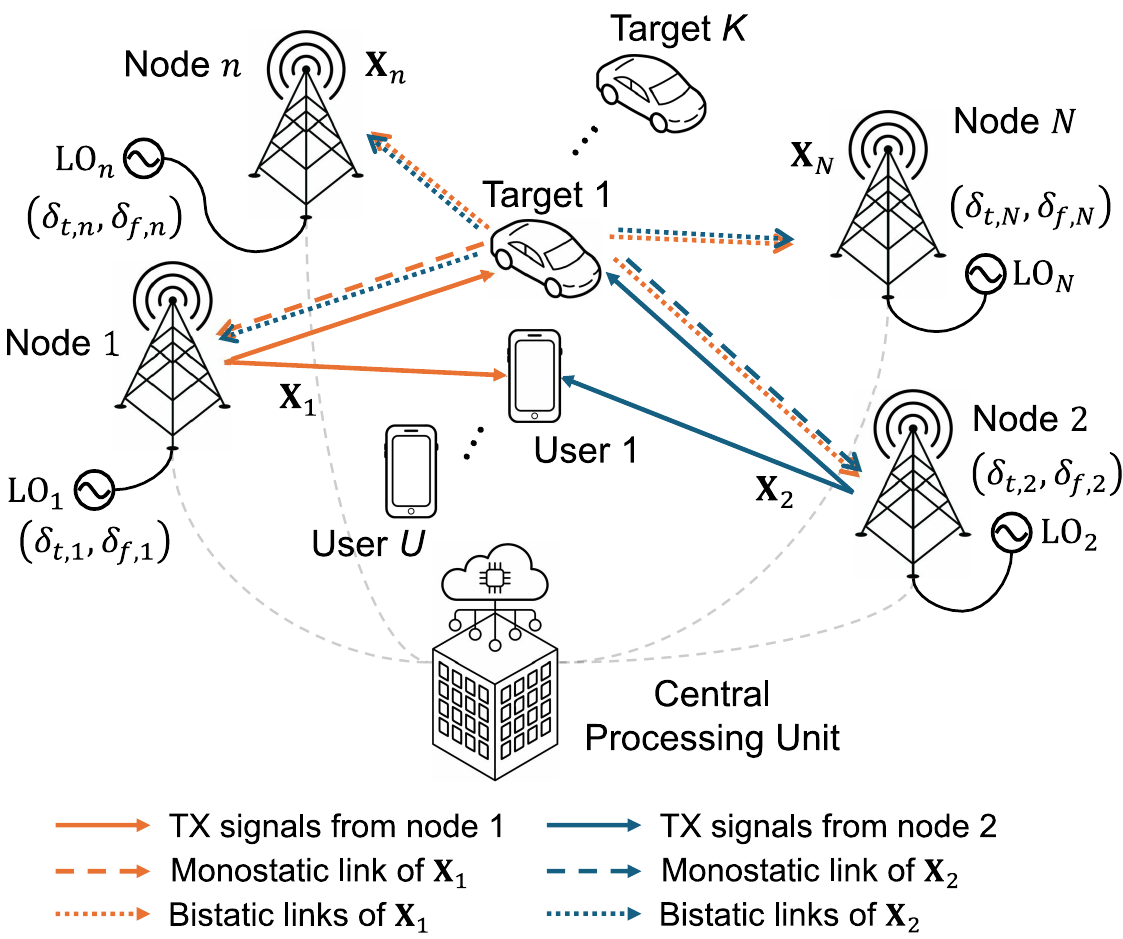}} 
    \caption{Illustration of the D-ISAC system with time and frequency offsets between nodes. Signals from other nodes are omitted for clarity. Each node is equipped with individual LO, resulting in time and frequency offsets between different nodes.}
    \label{f1}
\end{figure}

We consider a D-ISAC system, as shown in Fig. \ref{f1}, integrating cooperative communication with a distributed MIMO radar framework that utilizes both monostatic and bistatic sensing. By jointly exploiting monostatic and bistatic scattering for target localization, the system enhances sensing performance in ISAC networks while improving communication quality of service (QoS). To achieve this, all ISAC nodes are connected to a central processing unit (CPU) for data transfer, which distributes communication data to each node and collects radar-received signals or estimated target parameters. In the synchronization framework, we aim to achieve time and frequency synchronization across $N$ distributed ISAC nodes. 

We assume that the transmitter and receiver are separately deployed with sufficient isolation, sharing same local oscillator (LO) in each node. On the other hand, each node is equipped with individual LO, resulting in time and frequency offsets between different nodes. Additionally, we employ orthogonal frequency division multiplexing (OFDM) for both over-the-air synchronization and downlink ISAC operation. For clarity, the signal model is based on a general OFDM structure, encompassing preambles, pilots, and ISAC signals, including modulated communication data for transmission.

\subsection{Signal Model for Over-the-Air Synchronization}
The transmitted OFDM signal at the $n^{th}$ ISAC node in the D-ISAC system is given by  
\begin{align}
    s_n (t) = \frac{1}{\sqrt{P}}\sum_{q=0}^{Q-1} \sum_{p=0}^{P-1} X_n [p,q] e^{j(2\pi(f_c + \delta_{f,n} + p \Delta f)(t - \delta_{t,n}))}  \nonumber \\
    \times \; \Pi \left(\frac{t - \delta_{t,n} - qT}{T}\right),
    \label{Eq1}
\end{align}  
where $P$ and $Q$ represent the number of OFDM subcarriers and symbols, respectively. The term $X_n [p,q]$ denotes the $(p,q)$-th element of the matrix of modulated symbols $\mathbf{X}_n$ with normalized power $|X_n [p,q]|^2 = 1$, where $p$ and $q$ denote the subcarrier and OFDM symbol index, respectively. The carrier frequency is denoted by $f_c$, while $\delta_{t,n}$ and $\delta_{f,n}$ represent TO and CFO relative, respectively, to a reference node. Without loss of generality, we set the reference node index as $n=1$, yielding $\delta_{t,1}=0$ and $\delta_{f,1} = 0$.  We assume that the transmitted sequences $\mathbf{X}_n, \forall{n}=1,2, ...,N$ are fully orthogonal, which can be achieved using code-division multiplexing (CDM) or frequency-division multiplexing (FDM). Additionally, a cyclic prefix (CP) is omitted in this formulation, as CP removal during receiver processing prevents inter-symbol interference (ISI) \cite{han2023sub}.

Since all distributed ISAC nodes operate in the same time-frequency domain, the received signal at each ISAC node is the summation of all reflected signals from $K$ scatters. The received baseband signal at the $n^{th}$ ISAC node after down-conversion is given by  
\begin{align}
    y_n (t) = \sum_{m=1}^{N} y_{n,m} (t) + z_n (t),
    \label{Eq2}
\end{align}
where $z_n (t)$ represents zero-mean complex-valued additive white Gaussian noise (AWGN) at receiver $n$, following $z_n(t) \sim \mathcal{CN}(0,\sigma^2)$. The term $y_{n,m} (t)$ denotes the received signal at node $n$, which is transmitted from the $m^{th}$ node and reflected by scatters. This can be explicitly expressed as  
\begin{align}
    y_{n,m} (t) = \frac{1}{\sqrt{P}} \sum_{q=0}^{Q-1} \sum_{p=0}^{P-1} X_m [p,q] e^{j2\pi(\Delta \delta_{f,n,m} + p \Delta f) t} \nonumber \\
    \times \sum_{k=0}^{K-1} \beta_{n,m}^k e^{-j(2\pi(p \Delta f(\tau_{n,m}^k + \Delta \delta_{t,n,m}) - qT(f_{d,n,m}^k + \Delta \delta_{f,n,m}))}  \nonumber \\
    \times \Pi \left(\frac{t - \tau_{n,m}^k - \Delta \delta_{t,n,m} - qT}{T}\right),
    \label{Eq3}
\end{align}
where $\Delta \delta_{t,n,m}  = \delta_{t,m} - \delta_{t,n}$ and $\Delta \delta_{f,n,m}  = \delta_{f,m} - \delta_{f,n}$, representing the relative TO and CFO between node $n$ and node $m$. The term $\beta_{n,m}^k$ represents the complex amplitude of the $k^{th}$ target, incorporating the radar cross-section (RCS) and path loss associated with the target-to-node distance. The parameters $\tau_{n,m}^k$ and $f_{d,n,m}^k$ denote the time-of-flight (TOF) and Doppler frequency shift, respectively, for the signal traveling from the $m^{th}$ node to the $k^{th}$ target and then to the $n^{th}$ node.

By performing FFT on each symbol, (\ref{Eq2}) can be represented in the subcarrier-symbol domain as  
\begin{align}
    \mathbf{Y}_n  = \sum_{m=1}^{N} \mathbf{Y}_{n,m}  +\mathbf{Z}_n,
    \label{Eq6}
\end{align}
where $\mathbf{Y}_{n,m} \in \mathbb{C}^{P \times Q}$ and $\mathbf{Z}_n \in \mathbb{C}^{P \times Q}$ are the subcarrier-symbol domain matrix of $y_{n,m} (t)$ and $z_n (t)$, respectively. To simplify the expression, we assume $f_{d,n,m}^k + \Delta \delta_{f,n,m} \ll \Delta f/10$, ensuring that the OFDM system remains free from inter-carrier interference (ICI). Under this assumption, $\mathbf{Y}_{n,m}$ can be expressed as  
\begin{align}
    \mathbf{Y}_{n,m}  & = {\mathbf{H}}_{n,m} \odot \mathbf{X}_{m},
\end{align}
where the sensing channel matrix ${\mathbf{H}}_{n,m} \in \mathbb{C}^{P \times Q}$ is given by
\begin{align}
    {\mathbf{H}}_{n,m}  & = (\mathbf{\Lambda}_{t,n,m} \mathbf{R}_{n,m})\mathbf{B}_{n,m} (\mathbf{\Lambda}_{f,n,m} \mathbf{D}_{n,m})^T \nonumber \\
                        & = \mathbf{\Lambda}_{t,n,m} (\mathbf{R}_{n,m}\mathbf{B}_{n,m} \mathbf{D}_{n,m}) \mathbf{\Lambda}_{f,n,m} \nonumber \\
                        & = \mathbf{\Lambda}_{t,n,m} \Psi_{n,m} \mathbf{\Lambda}_{f,n,m}.
    \label{Eq8}
\end{align}
Here, $\Psi_{n,m} = \mathbf{R}_{n,m}\mathbf{B}_{n,m} \mathbf{D}_{n,m} \in \mathbb{C}^{P \times Q}$ represents the true target channel matrix, free from TO and CFO effects. The matrix $\mathbf{B}_{n,m} \in \mathbb{C}^{K \times K}$ contains the complex amplitudes, incorporating radar cross-section (RCS) and path loss. The matrices $\mathbf{R}_{n,m} \in \mathbb{C}^{P \times K}$ and $\mathbf{D}_{n,m} \in \mathbb{C}^{Q \times K}$ represent the delay and Doppler manifold matrices, respectively. Additionally, $\mathbf{\Lambda}_{t,n,m} \in \mathbb{C}^{P \times P}$ and $\mathbf{\Lambda}_{f,n,m} \in \mathbb{C}^{Q \times Q}$ are diagonal matrices characterizing TO and CFO. Specifically, they are defined as
\begin{align}
    \textbf{R}_{n,m} & = 
    \begin{bmatrix}
        \textbf{r}(\tau_{n,m}^1), & \textbf{r}(\tau_{n,m}^2), & \cdots & ,\textbf{r}(\tau_{n,m}^K)
    \end{bmatrix}, \\
    \textbf{D}_{n,m} & = 
    \begin{bmatrix}
        \textbf{d}(f_{d,n,m}^1), & \textbf{d}(f_{d,n,m}^2), & \cdots & , \textbf{d}(f_{d,n,m}^K)
    \end{bmatrix}, \\
    \textbf{B}_{n,m} & = 
    \text{diag}(\beta_{n,m}^1,\beta_{n,m}^2, \cdots, \beta_{n,m}^K), \\
    \mathbf{\Lambda}_{t,n,m} & = 
    \text{diag}(1, \cdots, e^{-j2 \pi (P-1) \Delta f \Delta \delta_{t,n,m}}), \label{TOstr} \\
    \mathbf{\Lambda}_{f,n,m} & = 
    \text{diag}(1, \cdots, e^{-j2 \pi (Q-1) T \Delta \delta_{f,n,m}}), \label{CFOstr}
\end{align}
where $\textbf{r}(\tau) = \begin{bmatrix} 1, e^{-j2 \pi \Delta f \tau}, \cdots, e^{-j2 \pi (P-1)\Delta f \tau} \end{bmatrix}^T$ and $\textbf{d}(f_d) = \begin{bmatrix} 1, e^{j 2 \pi T f_d},  \cdots, e^{j 2 \pi (Q-1) T f_d} \end{bmatrix}^T.$
In (\ref{Eq8}), it is readily observed that the target channel $\mathbf{H}_{n,m}$ includes the effects of TO $\mathbf{\Lambda}_{t,n,m}$ and CFO $\mathbf{\Lambda}_{f,n,m}$ between nodes $n$ and $m$. These effects introduce distortions in target range and Doppler estimation unless properly compensated.

\subsection{Sensing Channel Demodulation}
To estimate a specific sensing channel ${\mathbf{H}}_{n,m}$ from $\mathbf{Y}_{n}$, we perform frequency-domain matched filtering at node $n$ using the transmitted signal from node $m$. Let $\widehat{\mathbf{H}}_{n,m}$ denote the estimated sensing channel, which can be expressed as  
\begin{align}
    \widehat{\mathbf{H}}_{n,m} & = \mathbf{Y}_n \odot \mathbf{X}_m^{*} \nonumber \\ 
    & = \mathbf{H}_{n,m} + \sum_{i=1, i \neq m}^{N} \mathbf{H}_{n,i} \odot (\mathbf{X}_i \odot \mathbf{X}_m^{*}) + \overline{\mathbf{Z}}_n.
    \label{Eq7}
\end{align}
The second term represents the interference from other transmitting nodes when demodulating the sensing channel ${\mathbf{H}}_{n,m}$. Based on the assumed orthogonality of transmitted signals, this interference can be eliminated using the orthogonality condition $\mathbf{X}_i \odot \mathbf{X}_j^{*} = \mathbf{0}_{P \times Q}, \forall{i \neq j}$. Consequently, the demodulated sensing channel simplifies to  
\begin{align}
    \widehat{\mathbf{H}}_{n,m} & = \mathbf{H}_{n,m} + \overline{\mathbf{Z}}_n,
\end{align}
where $\overline{\mathbf{Z}}_n$ is the modified AWGN, following the same statistical distribution as ${\mathbf{Z}}_n$. In cases where transmitted signals are not perfectly orthogonal, additional processing may be required to suppress residual interference in MIMO radar \cite{wang2020slow}.

\section{Pairwise Synchronization of Distributed ISAC Using Reciprocal Bistatic Sensing Channels}
\subsection{Offset Reciprocity in Bistatic Observations}

\begin{figure}[t!]
    \centering
    \subfloat[]{\includegraphics[scale=0.55]{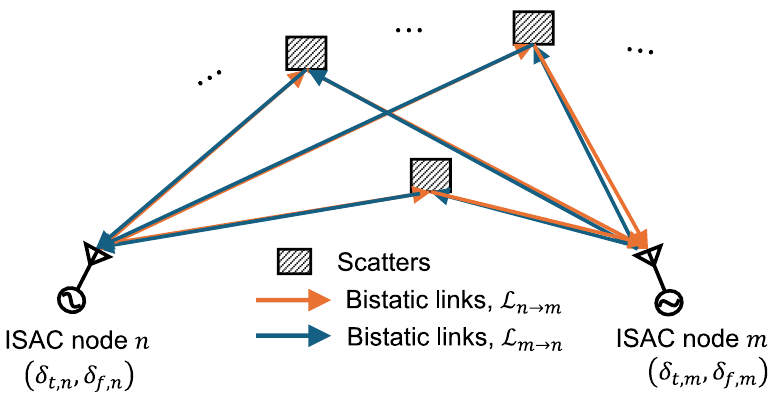}} \\
    \centering
    \subfloat[]{\includegraphics[scale=0.525]{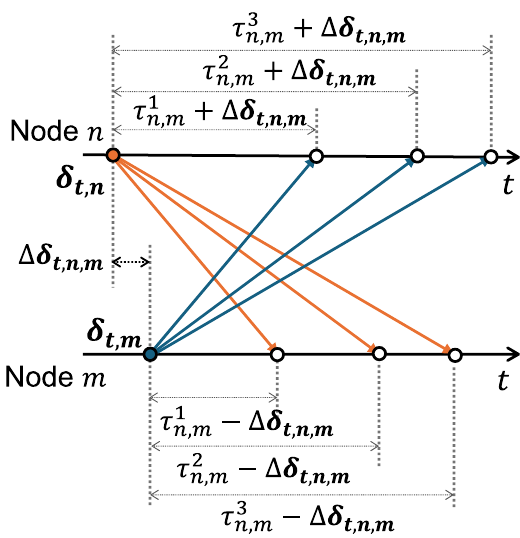}}
    \centering
    \subfloat[]{\includegraphics[scale=0.525]{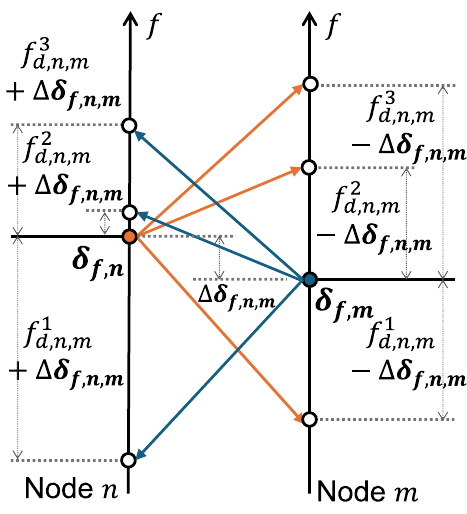}} 
    \caption{(a) Illustration of reciprocal bistatic sensing links utilized for the proposed over-the-air time-frequency synchronization. Relative (b) timing and (c) frequency diagrams between nodes are described with three scatters used for the synchronization.}
    \label{f2}
\end{figure}

In the proposed over-the-air synchronization framework, we exploit offset reciprocity in reciprocal bistatic sensing channels. Given the presence of several unknown reciprocal bistatic sensing links, as shown in Fig. \ref{f2}, these links can be leveraged to estimate TO and CFO between two ISAC nodes. Let $\mathcal{L}_{n\rightarrow m}$ denote the target sensing links from transmitter $n$ to the targets and then to receiver $m$. The key idea is that the true target range and Doppler parameters in $\mathcal{L}_{n\rightarrow m}$ and $\mathcal{L}_{m\rightarrow n}$ remain identical, while the effects of TO and CFO manifest symmetrically in opposite directions as shown in Fig. \ref{f2}(b) and \ref{f2}(c). By analyzing the sensing channel in (\ref{Eq8}), which consists of the true target channel and the offset between two nodes, the following proposition holds for reciprocal bistatic links:
\begin{proposition} \label{prop0}
The TO and CFO between node $n$ and node $m$ in the reciprocal bistatic sensing channels $\widehat{\mathbf{H}}_{n,m}$ and $\widehat{\mathbf{H}}_{m,n}$ satisfy the offset reciprocity as  
\begin{align}
    \mathbf{\Lambda}_{t,n,m} = \mathbf{\Lambda}_{t,m,n}^{*}, \\
    \mathbf{\Lambda}_{f,n,m} = \mathbf{\Lambda}_{f,m,n}^{*}.
\end{align}
\end{proposition}
\renewcommand\qedsymbol{$\blacksquare$}
\begin{proof}
The relative TO and CFO are defined as $\Delta \delta_{t,n,m}  = \delta_{t,m} - \delta_{t,n}$ and $\Delta \delta_{f,n,m}  = \delta_{f,m} - \delta_{f,n}$, which yield $\Delta \delta_{t,n,m}  = -\Delta \delta_{t,m,n}$ and $\Delta \delta_{f,n,m}  = -\Delta \delta_{f,m,n}$.
Substituting these into (\ref{TOstr}) and (\ref{CFOstr}) completes the proof.
\end{proof}
Based on our observation, the proved OR property in Proposition 1 can be leveraged to estimate TO and CFO between two nodes in the following sections. The proposed signal processing framework enables the use of this property without directly estimating target parameters. Instead, it focuses solely on estimating TO and CFO, irrespective of the number of targets and without requiring prior knowledge of targets or ISAC node locations, of which processing is developed in Section \ref{pre} and \ref{BSM}.

\subsection{Pre-processing for Delay-Doppler Decoupling} \label{pre}
Joint estimation of TO and CFO requires 2-D signal processing, which significantly increases computational complexity compared to 1-D processing. To mitigate this, we decouple delay and Doppler parameters before estimating TO and CFO between two ISAC nodes. This approach allows for independent estimation of TO and CFO through separate processing steps, reducing the overall complexity compared to 2-D joint estimation.

To this end, we first perform $P$-point inverse DFTs (IDFTs) along the subcarrier axis and $Q$-point DFTs along the symbol axis over $\widehat{\mathbf{H}}_{n,m}$. Let $\mathbf{C}_{n,m} = \mathbf{W}_P^{-1}\hat{\mathbf{H}}_{n,m}\mathbf{W}_Q$ denote the delay-Doppler spectrum, where $\mathbf{W}_{P}$ and $\mathbf{W}_{Q}$ denote the $P$-point DFT and $Q$-point DFT matrix, respectively. The delay and Doppler indices corresponding to the maximum spectral power can then be found as  
\begin{align}
    \hat{p}_{n,m} & = \arg \max_{p} \sum_{q = 0}^{Q-1} \left|C_{n,m}[p,q]\right|^2, \\
    \hat{q}_{n,m} & = \arg \max_{q} \sum_{p = 0}^{P-1} \left|C_{n,m}[p,q]\right|^2.
\end{align}
After identifying the indices of the maximal power column and row, $\mathbf{c}_{t,n,m} = C_{n,m}[p,\hat{q}_{n,m}]$ and $\mathbf{c}_{f,n,m}=C_{n,m}[\hat{p}_{n,m},q]$ are transformed into the Doppler-compressed sensing channel $\hat{\mathbf{h}}_{t,n,m} \in \mathbb{C}^{P \times 1}$ and the delay-compressed sensing channel $\hat{\mathbf{h}}_{d,n,m}^T \in \mathbb{C}^{1 \times Q}$ by performing DFT and IDFT, respectively, which can be written as
\begin{align}
    \hat{\mathbf{h}}_{t,n,m} = \mathbf{W}_{P} \mathbf{c}_{t,n,m}, \\
    \hat{\mathbf{h}}_{f,n,m} = \mathbf{W}_{Q}^{-1} \mathbf{c}_{f,n,m}^T.
\end{align}

Assuming that $K_t$ and $K_f$ targets are included in $\hat{\mathbf{h}}_{t,n,m}$ and $\hat{\mathbf{h}}_{d,n,m}$, respectively, where $K_t \leq K$ and $K_f \leq K$, they can be expressed as  
\begin{align}
    \hat{\mathbf{h}}_{t,n,m} & = \mathbf{\Lambda}_{t,n,m} {\psi}_{t,n,m} + \bar{\mathbf{z}}_{t,n,m},  \\
    \hat{\mathbf{h}}_{f,n,m} & = \mathbf{\Lambda}_{f,n,m} {\psi}_{f,n,m} + \bar{\mathbf{z}}_{f,n,m},
\end{align}
where  
\begin{align}
    {\psi}_{t,n,m} & = \mathbf{R}_{K_t,n,m}\mathbf{B}_{K_t,n,m} (Q \cdot \mathbf{1}_{1 \times K_t})^T, \label{Eqn::22} \\
    {\psi}_{f,n,m} & = \mathbf{D}_{K_f,n,m}^T\mathbf{B}_{K_f,n,m} (P \cdot \mathbf{1}_{1 \times K_f})^T. 
    \label{Eqn::23}
\end{align}
The terms $\bar{\mathbf{z}}_{t,n,m}$ and $\bar{\mathbf{z}}_{f,n,m}$ represent white Gaussian noise, following the distributions $\bar{\mathbf{z}}_{t,n,m}  \sim \mathcal{CN}(0,\sigma^2 Q \cdot \mathbf{I}_{P})$ and $\bar{\mathbf{z}}_{f,n,m} \sim \mathcal{CN}(0, \sigma^2 P \cdot \mathbf{I}_{Q})$. This pre-processing of delay and Doppler channels enables TO and CFO estimation to be performed as a 1-D process in the subsequent bistatic signal matching procedure.

\subsection{Bistatic Signal Matching} \label{BSM}
Before applying super-resolution estimators for TO and CFO estimation, we develop a bistatic signal matching (BSM) procedure to aggregate multiple scattered signals into a single offset for estimation. BSM enables fine-grained offset estimation using super-resolution algorithms without requiring prior knowledge of the number of sources. 

To estimate TO between the $n^{th}$ and $m^{th}$ ISAC nodes, we construct a matched signal $\mathbf{g}_{t,n,m}$ using the reciprocal bistatic sensing channels $\hat{\mathbf{h}}_{t,n,m}$ and $\hat{\mathbf{h}}_{t,m,n}$:
\begin{align}
    \mathbf{g}_{t,n,m}  & = \hat{\mathbf{h}}_{t,n,m} \odot \hat{\mathbf{h}}_{t,m,n}^{*}.
\end{align}
Leveraging the OR property in bistatic observations, this expression simplifies to  
\begin{align}
    \mathbf{g}_{t,n,m} & = \left(\mathbf{\Lambda}_{t,n,m} {\psi}_{t,n,m} + \bar{\mathbf{z}}_{t,n}\right) \odot \left(\mathbf{\Lambda}_{t,m,n} {\psi}_{t,m,n} + \bar{\mathbf{z}}_{t,m}\right)^{*}  \nonumber \\
    & = \mathbf{\Lambda}_{t,n,m}^2 ({\psi}_{t,n,m} \odot {\psi}_{t,n,m}^{*}) + \mathbf{\Lambda}_{t,n,m} \psi_{t,n,m} \odot \bar{\mathbf{z}}_{t,m}^{*} \nonumber \\ 
    & \quad \quad + \bar{\mathbf{z}}_{t,n} \odot \mathbf{\Lambda}_{t,n,m} \psi_{t,n,m}^{*} + \bar{\mathbf{z}}_{t,n} \odot \bar{\mathbf{z}}_{t,m}^{*}  \nonumber \\
    & = \mathbf{\Lambda}_{t,n,m}^2 ({\psi}_{t,n,m} \odot {\psi}_{t,n,m}^{*}) + \bar{\mathbf{z}}_{\Sigma,t,m,n}, \label{Eqn::25}
\end{align}
where $\bar{\mathbf{z}}_{\Sigma,t,m,n}$ represents the aggregate noise terms in $\mathbf{g}_{t,n,m}$.

By substituting (\ref{Eqn::22}) into (\ref{Eqn::25}), the first term can be further expanded as follows:
\begin{align}
    & \mathbf{\Lambda}_{t,n,m}^2 ({\psi}_{t,n,m} \odot {\psi}_{t,n,m}^{*})  \nonumber \\ 
    & = \mathbf{\Lambda}_{t,n,m}^2 Q^2 \left(\mathbf{R}_{K_t,n,m} \mathbf{B}_{K_t,n,m} \odot \mathbf{R}_{K_t,n,m}^{*} \mathbf{B}_{K_t,n,m}^{*}\right) \nonumber \\
    & = \mathbf{\Lambda}_{t,n,m}^2 Q^2 \left(\sum_{k=1}^{K_t} \beta_{n,m}^k \mathbf{r}(\tau_{n,m}^k)\right) \odot \left(\sum_{k=1}^{K_t} \left(\beta_{n,m}^k\right)^{*} \mathbf{r}(\tau_{n,m}^k)^{*}\right) \nonumber \\
    & = \left(Q^2 \sum_{k=1}^{K_t}(\beta_{n,m}^k)^2\right) \mathbf{r}\left(2 \Delta \delta_{t,n,m}\right) + \nonumber \\ 
    & Q^2 \sum_{i=1}^{K_t}\sum_{j=1, j \neq i}^{K_t} \beta_{n,m}^i \left(\beta_{n,m}^j\right)^{*} \mathbf{r}\left(\tau_{n,m}^i - \tau_{n,m}^j + 2 \Delta \delta_{t,n,m}\right). \label{Eqn::26} 
\end{align}
The first term in (\ref{Eqn::26}) represents the TO between nodes, where the power of all $K_t$ scatters is summed together. The second term, on the other hand, accounts for inter-modulated components arising from scatters located at different ranges.

Similarly, the matched signal $\mathbf{g}_{f,n,m}$ of delay-compressed reciprocal channels for CFO estimation is given by
\begin{align}
    \mathbf{g}_{f,n,m}  & = \hat{\mathbf{h}}_{f,n,m} \odot \hat{\mathbf{h}}_{f,m,n}^{*} \nonumber \\
    & = \left(\mathbf{\Lambda}_{f,n,m} {\psi}_{f,n,m} + \bar{\mathbf{z}}_{f,n}\right) \odot \left(\mathbf{\Lambda}_{f,m,n} {\psi}_{f,m,n} + \bar{\mathbf{z}}_{f,m}\right)^{*}  \nonumber \\
    & = \mathbf{\Lambda}_{f,n,m}^2 ({\psi}_{f,n,m} \odot {\psi}_{f,n,m}^{*}) + \mathbf{\Lambda}_{f,n,m} \psi_{f,n,m} \odot \bar{\mathbf{z}}_{f,m}^{*} \nonumber \\ 
    & \quad \quad + \bar{\mathbf{z}}_{f,n} \odot \mathbf{\Lambda}_{f,n,m} \psi_{f,n,m}^{*} + \bar{\mathbf{z}}_{f,n} \odot \bar{\mathbf{z}}_{f,m}^{*}  \nonumber \\
    & = \mathbf{\Lambda}_{f,n,m}^2 ({\psi}_{f,n,m} \odot {\psi}_{f,n,m}^{*}) + \bar{\mathbf{z}}_{\Sigma,f,m,n}, \label{Eqn::27}
\end{align}
where $\bar{\mathbf{z}}_{\Sigma,f,m,n}$ represents the aggregate noise terms in $\mathbf{g}_{f,n,m}$. Analogous to (\ref{Eqn::26}), the first term in (\ref{Eqn::27}) expands as
\begin{align}
    & \mathbf{\Lambda}_{f,n,m}^2 ({\psi}_{f,n,m} \odot {\psi}_{f,n,m}^{*})  = \left(P^2 \sum_{k=1}^{K_f}(\beta_{n,m}^k)^2\right) \mathbf{d}\left(2 \Delta \delta_{f,n,m}\right) \nonumber \\ 
    &  + P^2 \sum_{i=1}^{K_f}\sum_{j=1, j \neq i}^{K_f} \beta_{n,m}^i \left(\beta_{n,m}^j\right)^{*} \mathbf{d}\left(f_{d,n,m}^i - f_{d,n,m}^j + 2\Delta \delta_{f,n,m}\right). \label{Eqn::28}
\end{align}

\textbf{Remark 1:} The delay and Doppler vectors, $\mathbf{r}\left(2 \Delta \delta_{t,n,m}\right)$ and $\mathbf{d}\left(2 \Delta \delta_{f,n,m}\right)$, corresponding to TO and CFO, can be regarded as the principal components with the largest magnitude in the matched signals $\mathbf{g}_{t,n,m}$ and $\mathbf{g}_{f,n,m}$, respectively. This is because the by-products generated by inter-modulation between scatters are distributed across different delay and Doppler values, specifically at $\tau_{n,m}^i - \tau_{n,m}^j + 2 \Delta \delta_{t,n,m}$ and $f_{d,n,m}^i - f_{d,n,m}^j + 2\Delta \delta_{f,n,m}$, respectively.

Based on the observations in (\ref{Eqn::25}) and (\ref{Eqn::27}), the newly introduced noise terms $\bar{\mathbf{z}}_{\Sigma,t,m,n}$ and $\bar{\mathbf{z}}_{\Sigma,f,m,n}$ after the BSM processing have different noise distribution from the AWGN of the received signal. Thus, we characterize their statistical distribution in the following proposition.
\begin{proposition} \label{prop2}
Given that $\beta_{n,m}^k \sim \mathcal{CN}(0,|\beta_{n,m}^k|^2)$, $\forall k = 1, 2, \dots, K$, the aggregated noise components $\bar{\mathbf{z}}_{\Sigma,t,m,n}$ and $\bar{\mathbf{z}}_{\Sigma,f,m,n}$ follow a Gaussian distribution
\begin{align}
    \bar{\mathbf{z}}_{\Sigma,t,m,n} & \sim \mathcal{CN}(0, \sigma_t^2 \cdot \mathbf{I}_{P}), \\
    \bar{\mathbf{z}}_{\Sigma,f,m,n} & \sim \mathcal{CN}(0, \sigma_f^2 \cdot \mathbf{I}_{Q}),
\end{align}
where
\begin{align}
    \sigma_t^2 & = \sigma^2Q^2\left(\sum_{k=1}^{K_t} 2|\beta_{n,m}^k|^2 Q + \sigma^2\right), \\
    \sigma_f^2 & = \sigma^2P^2\left(\sum_{k=1}^{K_f} 2|\beta_{n,m}^k|^2 P + \sigma^2\right).
\end{align}
\end{proposition}

\renewcommand\qedsymbol{$\blacksquare$}
\begin{proof}
From (\ref{Eqn::25}), $\sigma_t^2\mathbf{I}_{P}$ is given by  
\begin{align}
    2 \mathbb{E}\left[\left(\mathbf{\Lambda}_{t,n,m} \psi_{t,n,m} \odot \bar{\mathbf{z}}_{t,m}^{*} \right)^2\right] 
    + \mathbb{E}\left[\left(\bar{\mathbf{z}}_{t,n} \odot \bar{\mathbf{z}}_{t,m}^{*} \right)^2\right] \nonumber \\
    = 2Q^2 \left(\sum_{k=1}^{K_t} |\beta_{n,m}^k|^2 Q \sigma^2 \mathbf{I}_{P} \right) + Q^2\sigma^4\mathbf{I}_{P}.
\end{align}
Thus, the expression for $\sigma_t^2$ is obtained as $\sigma_t^2 = \sigma^2Q^2\left(\sum_{k=1}^{K_t} 2|\beta_{n,m}^k|^2 Q + \sigma^2\right)$.
The expression for $\sigma_f^2$ can be derived using the same procedure.
\end{proof}
\noindent It turns out that the matched signal of the reciprocal bistatic sensing channels can be represented as a simple form of multiple exponential sources with added Gaussian noise. Consequently, TO and CFO estimation can be equivalently reformulated as the delay and Doppler estimation of the most significant source in the matched signals. This reformulation enables the application of simplified super-resolution methods for TO and CFO estimation.

\section{Super-resolution Offset Estimation for Distributed ISAC}
After the BSM processing described in Section \ref{BSM}, the TO and CFO between paired node have to be accurately estimated. To this end, we propose a super-resolution offset estimation (SOE) method based on the matched signal obtained through BSM. Owing to the BSM procedure, the design of the super-resolution offset estimator is simplified and made more computationally efficient. Hereafter, for notational convenience, we omit the node indices ${n,m}$. The matched signal can be rewritten as
\begin{align}
    \mathbf{g}_t & = \mathbf{\beta}_{\Sigma,t} \mathbf{r} (2\Delta \delta_{t}) + \sum_{k=1}^{\tilde{K}_{I,t}} \mathbf{\beta}_{I,t}^k \mathbf{r} \left(\tau_{I}^k + 2\Delta \delta_{t} \right) + \bar{\mathbf{z}}_{\Sigma,t}, \\
    \mathbf{g}_f & = \mathbf{\beta}_{\Sigma,f} \mathbf{d} (2\Delta \delta_{f}) + \sum_{k=1}^{\tilde{K}_{I,f}} \mathbf{\beta}_{I,f}^k \mathbf{d} \left(f_{d,I}^k + 2\Delta \delta_{f}\right) + \bar{\mathbf{z}}_{\Sigma,f},
\end{align}
where $\tilde{K}_{I,t} \leq K_t^2 - K_t$ and $\tilde{K}_{I,f} \leq K_f^2 - K_f$ represent the number of nuisance sources generated by inter-modulation. Additionally, the nuisance delay and Doppler terms are given by $\tau_{I}^k \in \{ \tau^i - \tau^j \mid i, j = 1, 2, \dots, K_t, \ i \neq j \}, \quad \forall k = 1, 2, \dots, \tilde{K}_{I,t}$ and $f_{d,I}^k  \in \{ f_d^i - f_d^j \mid i, j = 1, 2, \dots, K_f, \ i \neq j \}, \quad \forall k = 1, 2, \dots, \tilde{K}_{I,f}$.
Finally, $\mathbf{\beta}_{I,t}^k$ and $\mathbf{\beta}_{I,f}^k$ denote the corresponding complex amplitudes.

\subsection{Maximum Likelihood Estimation}
\begin{algorithm}[t!]
     \caption{SOE-MLE Using Reciprocal Bistatic Channels}
     \begin{algorithmic}[1]
     \renewcommand{\algorithmicrequire}{\textbf{Input:}}
     \renewcommand{\algorithmicensure}{\textbf{Output:}}
     \REQUIRE Estimated reciprocal bistatic sensing channels, $\widehat{\mathbf{H}}_{n,m}$ and $\widehat{\mathbf{H}}_{m,n}$
     \ENSURE TO and CFO between two ISAC nodes, $\Delta \widehat{\delta}_{t}$ and $\Delta \widehat{\delta}_{f}$ 
     
     \STATE Compute delay-Doppler spectra $\mathbf{C}_{n,m} = \mathbf{W}_P^{-1}\hat{\mathbf{H}}_{n,m}\mathbf{W}_Q$ and $\mathbf{C}_{m,n} = \mathbf{W}_P^{-1}\hat{\mathbf{H}}_{m,n}\mathbf{W}_Q$ from the estimated channels
     \STATE Extract maximal power indices $\hat{p}_{n,m}$ and $\hat{q}_{n,m}$ using (16) and (17)
     \STATE Transform $C_{n,m}[p,\hat{q}_{n,m}]$ and $C_{n,m}[\hat{p}_{n,m},q]$ into Doppler- and delay-compressed sensing channels, $\hat{\mathbf{h}}_{t,n,m}$ and $\hat{\mathbf{h}}_{d,n,m}$ using (18) and (19), respectively
     \STATE Compute $\mathbf{g}_{t}  = \hat{\mathbf{h}}_{t,n,m} \odot \hat{\mathbf{h}}_{t,m,n}^{*}$ and $\mathbf{g}_{f}  = \hat{\mathbf{h}}_{f,n,m} \odot \hat{\mathbf{h}}_{f,m,n}^{*}$
     \STATE Solve (\ref{ml1}) and (\ref{ml2}) with initial estimates:
        \[
        \Delta \widehat{\delta}_{t,0} = \frac{\hat{p}_{n,m} - \hat{p}_{m,n}}{2 P\Delta f}, \quad
        \Delta \widehat{\delta}_{f,0} = \frac{\hat{q}_{n,m} - \hat{q}_{m,n}}{2 QT}
        \]
     \end{algorithmic}
     \label{A1}
\end{algorithm}  

It is well known that the maximum likelihood (ML) estimator is the theoretically optimal unbiased estimator, capable of achieving the Cramér-Rao bound (CRB) under the condition of a sufficiently large number of samples. In this section, we present the ML estimation (MLE) of TO and CFO from the matched signal.
From (\ref{Eqn::25}) and (\ref{Eqn::27}), the negative log-likelihood functions of the matched signals $\mathbf{g}_t$ and $\mathbf{g}_f$ for all unknown parameters are given by
\begin{align}
    \mathcal{L}_t (\mathbf{g}_t) & = \left|\left|\mathbf{g}_t - \mathbf{\Lambda}_{t}^2 ({\psi}_{t} \odot {\psi}_{t}^{*}) \right|\right|^2 + c_t, \\
    \mathcal{L}_f (\mathbf{g}_f) & = \left|\left|\mathbf{g}_f - \mathbf{\Lambda}_{f}^2 ({\psi}_{f} \odot {\psi}_{f}^{*}) \right|\right|^2 + c_f,
\end{align}
where $c_t$ and $c_f$ are nuisance constants. We develop the following cost function to estimate TO and CFO:
\begin{align}
    F_t (\beta_t, 2\Delta \delta_{t}) & = \left|\left|\mathbf{g}_t - \beta_t \mathbf{r}(2\Delta \delta_{t}) \right|\right|^2, \label{cf1}\\
    F_f (\beta_f, 2\Delta \delta_{f}) & = \left|\left|\mathbf{g}_f - \beta_f \mathbf{d}(2\Delta \delta_{f}) \right|\right|^2. \label{cf2}
\end{align}
Unlike multiple-parameter estimation, the cost function depends only on a single source parameter because the parameters of interest are $\Delta \delta_{t}$ and $\Delta \delta_{f}$. The inter-modulation by-products $\tau_{I}^k$ and $f_{d,I}^k$ in (\ref{Eqn::26}) and (\ref{Eqn::28}) are nuisance parameters and can be ignored at the expense of the performance loss.

The TO and CFO estimates that minimize (\ref{cf1}) and (\ref{cf2}) are given by
\begin{align}
    (\widehat{\beta}_t, \Delta \widehat{\delta}_{t}) & =  \arg \min_{\beta_t, \Delta {\delta}_{t}} F_t (\beta_t, 2\Delta \delta_{t}), \label{ml1} \\
    (\widehat{\beta}_f, \Delta \widehat{\delta}_{f}) & =  \arg \min_{\beta_f, \Delta {\delta}_{f}} F_f (\beta_f, 2\Delta \delta_{f}). \label{ml2}
\end{align}
Since this formulation results in a nonlinear least squares (NLS) problem, which is computationally expensive, we adopt a two-step coarse-to-fine search approach to efficiently solve (\ref{ml1}) and (\ref{ml2}). The overall steps for the proposed SOE-MLE approach are summarized in Algorithm \ref{A1}.

\textbf{Remark 2:} Thanks to the pre-processing procedure for delay-Doppler decoupling in Section III-B, we can skip the coarse TO and CFO estimation step by leveraging the prior obtained information. The obtained indices via (18) and (19) are utilized to provide the initial estimate for the TO and CFO as $\Delta \widehat{\delta}_{t,0} = (\hat{p}_{n,m} - \hat{p}_{m,n})/(2 P\Delta f)$ and $\Delta \widehat{\delta}_{f,0} = (\hat{q}_{n,m} - \hat{q}_{m,n})/(2 QT)$. Then, the fine search for the problem can be conducted within the search boundary of $[\Delta \widehat{\delta}_{t,0} - 1/2 P\Delta f, \Delta \widehat{\delta}_{t,0} + 1/2 P\Delta f]$ and $[\Delta \widehat{\delta}_{f,0} - 1/2 QT, \Delta \widehat{\delta}_{t,0} + 1/2 QT]$. This can be implemented with a nonlinear programming solver.

\subsection{Matrix Pencil}
\begin{algorithm}[t!]
     \caption{SOE-MP Using Reciprocal Bistatic Channels}
     \begin{algorithmic}[1]
     \renewcommand{\algorithmicrequire}{\textbf{Input:}}
     \renewcommand{\algorithmicensure}{\textbf{Output:}}
     \REQUIRE Estimated reciprocal bistatic sensing channels, $\widehat{\mathbf{H}}_{n,m}$ and $\widehat{\mathbf{H}}_{m,n}$
     \ENSURE TO and CFO between two ISAC nodes, $\Delta \widehat{\delta}_{t}$ and $\Delta \widehat{\delta}_{f}$ 
     
     \STATE Same steps 1-4 of SOE-MLE in Algorithm \ref{A1}
     \STATE Perform SVD on the Hankel matrices $\mathcal{H}_{t}$ and $\mathcal{H}_{f}$
     \STATE Extract $\mathbf{v}_{t,1}$ and $\mathbf{v}_{t,2}$ as the first column vectors of the right singular matrix $\mathbf{V}_t$ after removing the last and first rows (similarly for $\mathbf{v}_{f,1}$ and $\mathbf{v}_{f,2}$)
     \STATE Compute TO and CFO using (47) and (48)
     \end{algorithmic}
     \label{A2}
\end{algorithm}

Although our proposed SOE-MLE method can achieve the optimal solution, its reliance on an exhaustive search results in high computational complexity. This makes it impractical when dealing with a large number of D-ISAC nodes. Therefore, as an alternative, we propose a suboptimal estimator based on the matrix pencil (MP) approach. The MP method begins by constructing a $(P-L_t) \times L_t$ Hankel matrix from the matched signal using a pencil parameter $L_t$, as follows:
\begin{align}
    \mathcal{H}_{t} & = 
    \begin{bmatrix}
        {g}_t[0] & {g}_t[1] & \cdots & {g}_t[L_t] \\
        {g}_t[1] & {g}_t[2] & \cdots & {g}_t[L_t+1] \\
        \vdots & \vdots & \ddots & \vdots \\
        {g}_t[P - L_t - 1] & {g}_t[P - L_t] & \cdots & {g}_t[P-1] \\
    \end{bmatrix}
\end{align}

The conventional MP approach exploits the shift-invariance property to extract the parameters of multiple exponential signals \cite{sarkar1995using}. It begins by performing singular-value decomposition (SVD) of the Hankel matrix $\mathcal{H}_{t}$ as
\begin{align}
    \mathcal{H}_{t} = \mathbf{U}_t \Sigma_t \mathbf{V}_t^H,
\end{align}
where $\mathbf{U}_t$ and $\mathbf{V}_t$ are unitary matrices whose columns are the left and right singular vectors of $\mathcal{H}_{t}$, respectively. The matrix $\Sigma_t$ is diagonal, containing the singular values of $\mathcal{H}_{t}$. Let $\mathbf{V}_{t,1}$ and $\mathbf{V}_{t,2}$ be sub-matrices of $\mathbf{V}_t$ obtained by deleting the last and first rows, respectively. Then, the eigenvalue problem is formulated as
\begin{align}
    \left(\mathbf{V}_{t,2}^H - \lambda \mathbf{V}_{t,1}^H\right) \mathbf{\nu} = 0, \label{Eqn::44}
\end{align}
where $\lambda$ represents the eigenvalue associated with the signal pole, and $\mathbf{\nu}$ is the corresponding eigenvector.

Interestingly, the MP procedure for SOE can be simplified due to the properties of the matched signal, similar to the case of SOE-MLE. We leverage the signal structure obtained through BSM, where the principal singular value and its corresponding singular vector are directly related to the offset parameter.

Let $\mathbf{v}_{t,1}$ and $\mathbf{v}_{t,2}$ be the vectors corresponding to the first column of $\mathbf{V}_{t,1}$ and $\mathbf{V}_{t,2}$, as they are the only eigenvectors related to the offset parameter. This processing also filters out unwanted noise, including inter-modulation effects between multiple targets in (\ref{Eqn::26}) and (\ref{Eqn::28}). The eigenvalue problem is then reformulated as
\begin{align}
    \left(\mathbf{v}_{t,2}^H - \lambda \mathbf{v}_{t,1}^H\right) \mathbf{\nu} = 0.
\end{align}
The solution to this equation can be obtained through a simple calculation:
\begin{align}
    \lambda = \frac{\mathbf{v}_{t,2}^H  \mathbf{v}_{t,1}}{\mathbf{v}_{t,1}^H \mathbf{v}_{t,1}},
\end{align}
where $\lambda = e^{-j2\pi \Delta f (2 \Delta \widehat{\delta}_{t})}$. Accordingly, the TO estimate for SOE-MP is given by
\begin{align}
    \Delta \widehat{\delta}_{t} = -\frac{1}{4\pi \Delta f} \tan^{-1} \left(\frac{\Im \left(\mathbf{v}_{t,2}^H  \mathbf{v}_{t,1}/\mathbf{v}_{t,1}^H \mathbf{v}_{t,1}\right)}{\Re\left(\mathbf{v}_{t,2}^H  \mathbf{v}_{t,1}/\mathbf{v}_{t,1}^H \mathbf{v}_{t,1}\right)}\right).
\end{align}
Following the same procedure as the TO estimation, the CFO estimate for SOE-MP is obtained as
\begin{align}
    \Delta \widehat{\delta}_{f} = \frac{1}{4\pi T} \tan^{-1} \left(\frac{\Im \left(\mathbf{v}_{f,2}^H  \mathbf{v}_{f,1}/\mathbf{v}_{f,1}^H \mathbf{v}_{f,1}\right)}{\Re\left(\mathbf{v}_{f,2}^H  \mathbf{v}_{f,1}/\mathbf{v}_{f,1}^H \mathbf{v}_{f,1}\right)}\right).
\end{align}
The overall steps for the proposed SOE-MP approach are summarized in Algorithm \ref{A2}.

\subsection{Cramer-Rao Bound for Offset Estimation}
To provide the theoretical performance bound for the proposed offset estimation, we derive the CRB expressions for TO and CFO estimation from the matched signals. The CRB results are presented in the following theorem.
\begin{theorem}\label{theo1}
Suppose that $\inf \{\tau_{I}^k\} \geq 1/P\Delta f$ or $\sup \{\beta_{I,t}^k\} \ll \beta_{\Sigma,t}$. Then, the CRB for the TO estimate is given by  
\begin{align}
    \textsc{CRB}_{\Delta \delta_{t}} = \frac{3 (1+2\gamma_t Q)}{8 \pi^2 (\Delta f)^2 P(P^2-1) \gamma_{t}^2 Q^2}, \label{Eqn::48}
\end{align}
where $\gamma_t = \sum_{k=1}^{K_t}|\beta_{n,m}^k|^2 / \sigma^2$.  

Similarly, if $\inf \{f_{d,I}^k\} \geq 1/QT$ or $\sup \{\beta_{I,f}^k\} \ll \beta_{\Sigma,f}$, the CRB for the CFO estimate is given by  
\begin{align}
    \textsc{CRB}_{\Delta \delta_{f}} = \frac{3 (1+2\gamma_f P)}{8 \pi^2 T^2 Q (Q^2-1) \gamma_{f}^2 P^2}, \label{Eqn::49}
\end{align}
where $\gamma_f = \sum_{k=1}^{K_f}|\beta_{n,m}^k|^2 / \sigma^2$.
\end{theorem}
\renewcommand\qedsymbol{$\blacksquare$}
\begin{proof}
It is straightforward to show that the CRB of the single-tone parameter estimate from the discrete signal with $P$ sample length and $T_s$ sampling period is given by \cite{rife1974single}
\begin{align}
    \textsc{CRB} = \frac{3}{2\pi^2\gamma T_s^2 P(P^2-1)}, \label{Eqn::Theo1}
\end{align}
where $\gamma$ represents the SNR of the signal. Based on the Proposition 2, we substitute $T_s = \Delta f$ and $\gamma = \frac{Q^2\gamma_t^2}{ 1+2\gamma_t Q}$ into (\ref{Eqn::Theo1}), yielding the CRB for the TO estimate from (\ref{Eqn::25}). Since the parameter to be estimated in $\mathbf{g}_t$ corresponds to $2\Delta \delta_t$, the CRB for $\Delta \delta_t$ estimate is reduced by the factor of 4, leading to (\ref{Eqn::48}). Similarly, substituting $T_s = T$ and $\gamma = \frac{P^2\gamma_f^2}{ 1+2\gamma_f P}$ into (\ref{Eqn::Theo1}) gives the CRB for the CFO estimate as (\ref{Eqn::49}).
\end{proof}
\textbf{Remark 3:} For a sufficiently large SNR, $\gamma_t$ and $\gamma_f$, it can be observed that the CRB for TO estimation is inversely proportional to $(P\Delta f)^2 P Q$, while the CRB for CFO estimation is inversely proportional to $(Q T)^2 P Q$. This implies that a large coherence bandwidth (CB) enhances TO estimation accuracy, whereas a long coherent processing interval (CPI) improves CFO estimation accuracy. Additionally, strong reflections from a large number of scatterers are desirable, as they increase the SNRs $\gamma_t$ and $\gamma_f$, leading to more accurate estimation.

\subsection{Centered Pairwise Synchronization of Distributed ISAC}
The pairwise synchronization technique employing the SOE approach is further extended to the general case with $N$ nodes in the D-ISAC system. To achieve synchronization across all nodes, SOE is performed for $(N-1)$ pairs, where each pair consists of a selected reference node and another node. This synchronization strategy is referred to as centered pairwise synchronization (CPS). In this section, we derive the theoretical bound for the expected synchronization performance of D-ISAC using stochastic geometry with the consideration of practical and random ISAC node locations.

For simplicity of analysis, we assume a single-target case. Let the node 1 be a reference node. Additionally, let $\sigma_{\Delta \delta_{t,n}}^2$ and $\sigma_{\Delta \delta_{f,n}}^2$ denote the estimate variances of the TO and CFO between node 1 and node $n$, where $n = 2, 3, \dots, N$. Then, the total offset estimate variance for CPS can be expressed as:
\begin{align}
    \widetilde{\sigma}_{\Delta \delta_{t}}^2 & = \sum_{n=2}^{N} \sigma_{\Delta \delta_{t,n}}^2 \geq \sum_{n=2}^{N} \textsc{CRB}_{\Delta \delta_{t,1,n}}, \label{Eqn::50}\\
    \widetilde{\sigma}_{\Delta \delta_{f}}^2 & = \sum_{n=2}^{N} \sigma_{\Delta \delta_{f,n}}^2 \geq \sum_{n=2}^{N} \textsc{CRB}_{\Delta \delta_{f,1,n}}, 
\end{align}
where the CRB for the offset estimate is given in Theorem \ref{theo1}.

Now, considering path loss, the received signal power from the target can be expressed as:
\begin{align}
    \left|\beta_{n,m}\right|^2 = \eta \alpha R_{n}^{-2} R_{m}^{-2}, \label{Eqn::52}
\end{align}
where $\eta$ is a gain factor incorporating the antenna gain and transmit power, and $\alpha$ represents the RCS of the target. The term $R_{n}$ denotes the distance between the $n^{th}$ node and the target.

To consider randomly deployed D-ISAC nodes, we assume that the node distribution follows a Poisson point process (PPP) with an intensity of $\mu$ \cite{meng2024bs}. Then, the distances $R_{1}, R_{2}, \dots , R_{N}$ are independent random variables following a Rayleigh distribution. The following lemma provides the optimal selection of the reference node for CPS in the D-ISAC system.

\begin{lemma} \label{lemma1}
Given that the distributed ISAC nodes are indexed in increasing order based on their distance from the target as $R_{1} \leq R_{2} \leq \dots \leq R_{N}$, the node closest to the target, i.e., $n = 1$, is the optimal choice as the reference node.
\end{lemma}

\renewcommand\qedsymbol{$\blacksquare$}
\begin{proof}
Please refer to Appendix \ref{proof_lemma}.
\end{proof}
Lemma \ref{lemma1} presents the optimal reference node selection for the CPS using reciprocal bistatic measurements in D-ISAC. Intuitively, this is because the pairwise offset estimation accuracy is proportional to the target SNR and the reference node closest to the target can provide the highest SNR for the reciprocal bistatic measurements with other nodes.
Based on Lemma \ref{lemma1}, the following theorem establishes a bound on the total offset estimate variance.
\begin{theorem} \label{theo2}
The total offset estimate variance for TO and CFO in a system with $N$ distributed ISAC nodes can be approximately bounded as
\begin{align}
    \widetilde{\sigma}_{\Delta \delta_{t}}^2 & \geq \frac{3 \sigma^2 (N-1) (N+2)}{8 \pi^4 \eta \alpha (\Delta f)^2  \mu^2 P^3 Q}, \\
    \widetilde{\sigma}_{\Delta \delta_{f}}^2 & \geq \frac{3 \sigma^2 (N-1) (N+2)}{8 \pi^4 \eta \alpha T^2  \mu^2 P Q^3}. \label{Eqn::56}
\end{align}
\end{theorem}

\renewcommand\qedsymbol{$\blacksquare$}
\begin{proof}
Please refer to Appendix \ref{proof2}.
\end{proof}

\textbf{Remark 4:} From Theorem \ref{theo2}, it can be readily inferred that the overall synchronization performance remains constant under a fixed area condition as the number of nodes increases. Conversely, under a fixed node density condition, the overall synchronization performance deteriorates as the number of nodes grows. The analysis of synchronization performance in D-ISAC will be further examined through numerical simulations in Section V.

\subsection{Target Localization Error in Asynchronous D-ISAC}
To analyze the impact of TOs on target localization error in D-ISAC, we derive the target localization CRB for asynchronous D-ISAC systems. Unlike conventional distributed MIMO radar, which relies on separately deployed transmitter and receiver nodes, the D-ISAC system features collocated transmitters and receivers at each node. Therefore, it leverages both monostatic and bistatic measurements for the target localization, introducing a new dependency of localization performance on the system's synchronization level.

Considering a target located at $[x, y]^T$ and D-ISAC node $n$ at $[x_n, y_n]^T$, the estimated TOF $\hat{\tau}_{n,m}$ from transmitter $m$ to the target and then to receiver $n$ for target localization is expressed as
\begin{align}
    \hat{\tau}_{n,m} & = \tau_{n,m} + \Delta {\delta}_{t,n,m} + z_{R,n,m}, \\
    \tau_{n,m} & = \frac{\sqrt{(x - x_n)^2 + (y - y_n)^2} + \sqrt{(x - x_m)^2 + (y - y_m)^2}}{c},
\end{align}
where $z_{R,n,m} \sim \mathcal{N}(0,\sigma_{R,n,m}^2)$ and $\sigma_{R,n,m}^2 = \frac{3 \sigma^2}{2 \pi^2 \left|\beta_{n,m}\right|^2 (\Delta f)^2  P^3 Q}$. Let $\mathbf{\Psi}$ denote the collection of all the real-valued unknown parameters $\mathbf{\Psi} = [\boldsymbol{\tau}^T, {\Delta \boldsymbol{\delta}}^T]^T$, where $\boldsymbol{\tau} = [\boldsymbol{\tau}_1^T, \boldsymbol{\tau}_2^T, \cdots, \boldsymbol{\tau}_N^T]^T,  \boldsymbol{\tau}_n = [\tau_{n,1}, \tau_{n,2}, \cdots, \tau_{n,N}]^T \in \mathbb{R}^{N \times 1}$ and $\boldsymbol{\Delta {\delta}} = [\boldsymbol{\Delta {\delta}}_1^T, \boldsymbol{\Delta {\delta}}_2^T, \cdots, \boldsymbol{\Delta {\delta}}_N^T]^T \in \mathbb{R}^{N^2-N \times 1}, \boldsymbol{\Delta {\delta}}_n = [\Delta {\delta}_{t,n,1}, \Delta {\delta}_{t,n,2}, \cdots, \Delta {\delta}_{t,n,N}]^T \in \mathbb{R}^{N-1 \times 1}$. Since no TO between TX and RX within a node is assumed, we do not need to consider $\Delta {\delta}_{t,n,n}, \forall n=1,2,... N$. 

Since the time synchronization error is an unknown random parameter, we derive a hybrid CRB (HCRB) using the FIM, which has the following block structure \cite{song2021target}:
\begin{align}
\mathbf{F}(\mathbf{\Psi}) & = \mathbf{F}_D(\mathbf{\Psi}) + \mathbf{F}_P(\mathbf{\Psi}) \\
    & = \begin{bmatrix}
        \mathbf{F}_{\boldsymbol{\tau}\boldsymbol{\tau}} & \mathbf{F}_{\boldsymbol{\tau}\Delta \boldsymbol{\delta}} \\ 
        \mathbf{F}_{\Delta \boldsymbol{\delta}\boldsymbol{\tau}} &
        \mathbf{F}_{\Delta \boldsymbol{\delta}\Delta \boldsymbol{\delta}} + \mathbf{F}_{P, \Delta \boldsymbol{\delta}\Delta \boldsymbol{\delta}}
    \end{bmatrix},
    \label{Eqn::17}
\end{align}
where $\mathbf{F}_{P, \Delta \boldsymbol{\delta}\Delta \boldsymbol{\delta}}$ represents the FIM of the prior information on TOs. Since our objective is to evaluate target localization performance in Cartesian coordinates, we define the vector of unknown parameters as $\boldsymbol{\Theta} = [x, y, \boldsymbol{\Delta {\delta}}^T]^T$. 

Applying the chain rule, the FIM with respect to $\boldsymbol{\Theta}$ can be obtained as
\begin{align}
    \mathbf{F(\boldsymbol{\Theta})}  =   
    \begin{bmatrix}
        \mathbf{J}^T \mathbf{F}_{\boldsymbol{\tau}\boldsymbol{\tau}} \mathbf{J} & \mathbf{J}^T \mathbf{F}_{\boldsymbol{\tau}\Delta \boldsymbol{\delta}} \\ 
        \mathbf{F}_{\Delta \boldsymbol{\delta}\boldsymbol{\tau}}\mathbf{J} &
        \mathbf{F}_{\Delta \boldsymbol{\delta}\Delta \boldsymbol{\delta}} + \mathbf{F}_{P, \Delta \boldsymbol{\delta}\Delta \boldsymbol{\delta}}
    \end{bmatrix},
    \label{Eqn::15}
\end{align} 
where $\mathbf{J} = [\frac{\partial \boldsymbol{\tau}}{\partial {x}} \frac{\partial \boldsymbol{\tau}}{\partial {y}}]$ is the Jacobian matrix. Finally, we get the target localization $\text{CRB}_{L}$, which is expressed as
\begin{equation}
    \text{CRB}_{L} = \text{tr} \left(\left[\mathbf{J}^T \mathbf{F}_{\boldsymbol{\tau}\boldsymbol{\tau}} \mathbf{J} - \mathbf{J}^T \mathbf{F}_{\boldsymbol{\tau}\Delta \boldsymbol{\delta}} (\mathbf{F}_{\Delta \boldsymbol{\delta}\Delta \boldsymbol{\delta}} + \mathbf{F}_{P, \Delta \boldsymbol{\delta}\Delta \boldsymbol{\delta}})^{-1} \mathbf{F}_{\Delta \boldsymbol{\delta}\boldsymbol{\tau}}\mathbf{J} \right]^{-1} \right).
    \label{Eqn::64}
\end{equation}

\subsubsection{Centralized Processing}
In centralized processing, we leverage signal-level fusion from all ISAC nodes to localize the target. Consequently, $N$ monostatic links and $(N^2 - N)$ bistatic links contribute to the target localization. We exploit the symmetric property of bistatic observations, $\tau_{n,m} = \tau_{m,n}$, and the reciprocal TOs in bistatic links, $\Delta {\delta}_{t,n,m} = -\Delta {\delta}_{t,m,n}$, to reduce the number of unknown parameters in $\mathbf{\Psi}$. To this end, the unknown parameters are rearranged to eliminate redundancy as $\boldsymbol{\tau} = [\tau_{1,1}, \tau_{1,2}, \dots, \tau_{N,N}]^T \in \mathbb{R}^{\frac{N(N+1)}{2} \times 1}$, representing the TOFs, and $\boldsymbol{\Delta {\delta}} = [\Delta {\delta}_{t,1,2}, \dots, \Delta {\delta}_{t,N,N-1}]^T \in \mathbb{R}^{\frac{N(N-1)}{2} \times 1}$, representing the TO errors. Although the centralized processing outperforms the decentralized one, it requires high backhaul capacity to share raw signals from multiple nodes.

\subsubsection{Decentralized Processing}
For decentralized processing, target localization is performed independently at each node. In this case, only the estimated TOFs from one monostatic link and $(N-1)$ bistatic links are combined for localization. Thus, the unknown parameters $\boldsymbol{\tau}$ are modified to $\boldsymbol{\tau}_n = [\tau_{n,1}, \tau_{n,2}, \dots, \tau_{n,N}]^T \in \mathbb{R}^{N \times 1}$ when localizing the target at node $n$, which includes only the measurements associated with node $n$. Similarly, the TO parameter is given by $\boldsymbol{\Delta {\delta}} = [\Delta {\delta}_{t,n,1}, \dots, \Delta {\delta}_{t,n,N}]^T \in \mathbb{R}^{(N-1) \times 1}$, where $\Delta {\delta}_{t,n,n}$ is omitted. The estimated target location in each node's decentralized processing can be fused and averaged in the CPU.

\section{Numerical Simulations}
We present numerical simulation results for the proposed over-the-air time-frequency synchronization technique and D-ISAC performance analysis under synchronization errors. Unless stated otherwise, the overall system parameters are set as follows: the operating frequency is $f_c = 5$ GHz, the OFDM signal bandwidth is $B = 50$ MHz, and the number of OFDM subcarriers and symbols are $P = 64$ and $Q = 32$, respectively. Since the proposed method is designed to provide fine-grained offset estimation and compensation, the standard deviations of the initial TO and CFO are set to $20 \; n$s and $10$ kHz, respectively.

\subsection{Performance of Super-resolution Offset Estimation}
\begin{figure}[t!]
    \centering
    \subfloat[]{\includegraphics[scale=0.575]{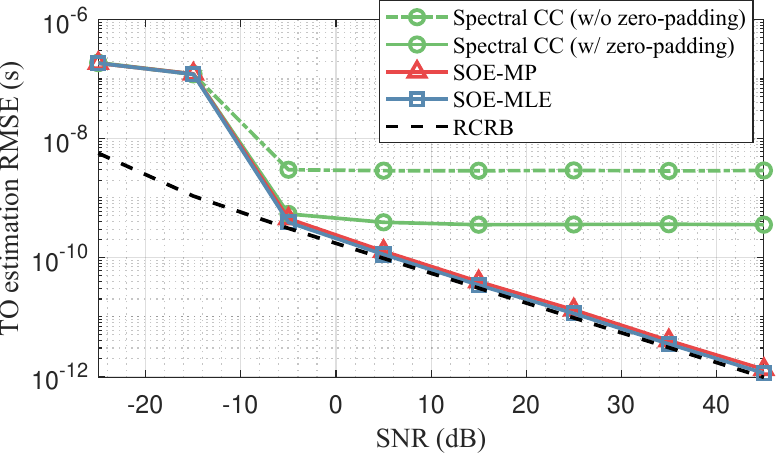}}  \\
    \centering
    \subfloat[]{\includegraphics[scale=0.575]{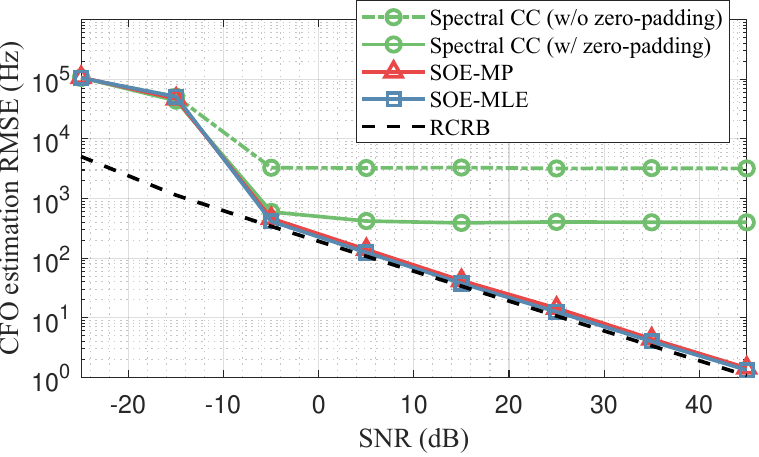}} 
    \caption{Time-frequency offset estimation performance between two ISAC nodes using reciprocal bistatic sensing channels: (a) TO estimation, and (b) CFO estimation.}
    \label{f3}
\end{figure}

We first investigate the performance of the proposed over-the-air time-frequency synchronization technique. For performance evaluation, we compare the proposed SOE techniques with a benchmark method based on spectral cross-correlation (CC). The spectral CC approach refers to a method that cross-correlates two bistatic measurement spectra, $\mathbf{C}_{n,m}$ and $\mathbf{C}_{m,n}$, as described in Section III-B. Due to the limited accuracy of the spectral CC method caused by its on-grid property, we further increased the number of grid points by applying zero-padding, providing eight times finer grid resolution compared to the case without zero-padding. In this simulation, we assume a single point scatter ($K = 1$) and consider two ISAC nodes positioned at \([35.35 \, \mathrm{m}, -35.35 \, \mathrm{m}]\) and \([-35.35 \, \mathrm{m}, -35.35 \, \mathrm{m}]\), respectively. Each result is obtained by 1000 runs of Monte Carlo simulation.

The TO and CFO estimation performances using the reciprocal bistatic channels are demonstrated in Fig. \ref{f3}. It is clearly observed that the spectral CC-based offset estimation exhibits accuracy limitations in both TO and CFO estimation, even for a high-SNR target. By increasing the number of grid points through zero-padding, spectral CC improves estimation accuracy compared to the case without zero-padding. However, it remains insufficient for fine-grained synchronization and suffers from an error floor, achieving approximately 500 ps and 500 Hz accuracy for TO and CFO estimation, respectively. In contrast, the proposed SOE-MLE and SOE-MP techniques provide significantly improved offset estimation performance, which scales proportionally with target SNR. Moreover, their performance follows the root CRB (RCRB) derived in Theorem \ref{theo1} when the target SNR is sufficiently high. Although the low-complexity SOE-MP estimator exhibits slightly degraded performance compared to SOE-MLE, the performance gap remains negligible. These results indicate that the proposed technique enables fine-grained time-frequency synchronization when strong target reflections are present. Consequently, robust synchronization in D-ISAC systems can be achieved by leveraging strong scatterers such as concrete buildings and large landmarks in urban environments.

\begin{figure}[t!]
    \centering
    \subfloat[]{\includegraphics[scale=0.6]{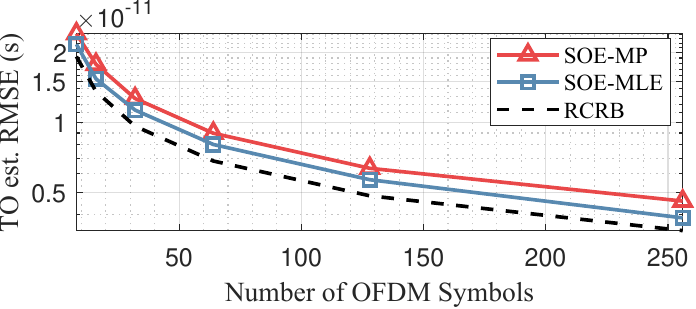}}  \\
    \centering
    \subfloat[]{\includegraphics[scale=0.6]{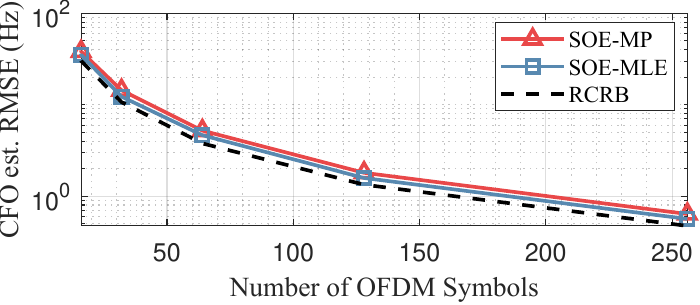}} 
    \caption{Impact of the number of OFDM symbols on the proposed SOE approaches: (a) TO estimation, and (b) CFO estimation.}
    \label{f4}
\end{figure}

\begin{figure}[t!]
    \centering
    \subfloat[]{\includegraphics[scale=0.6]{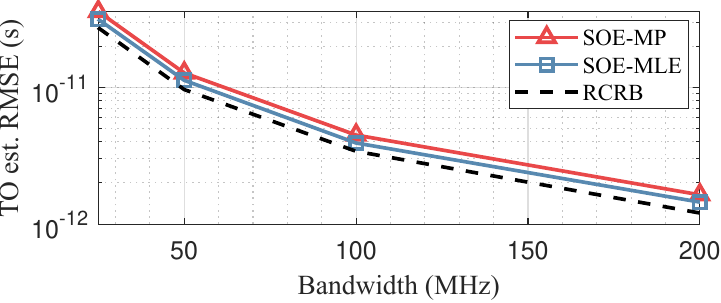}}  \\
    \centering
    \subfloat[]{\includegraphics[scale=0.6]{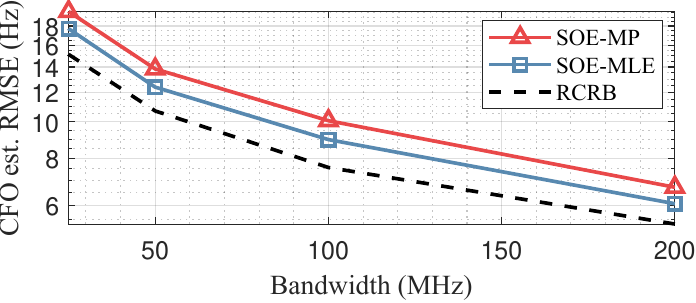}} 
    \caption{Impact of OFDM signal bandwidth on the proposed SOE approaches: (a) TO estimation, and (b) CFO estimation.}
    \label{f5}
\end{figure}

Next, we investigate the impact of system parameters on the proposed SOE performance. In this simulation, the target SNR is fixed at 25 dB. Fig. \ref{f4} illustrates the effect of the number of OFDM symbols, $Q$, used for synchronization on TO and CFO estimation. Based on the theoretical analysis in Theorem \ref{theo1}, it is expected that both TO and CFO estimation improve as more OFDM symbols are utilized. This is because a longer CPI provides increased processing gain, enhancing offset estimation accuracy in BSM signals. The contribution of signal bandwidth is also analyzed in Fig. \ref{f5}. In this simulation, a fixed subcarrier spacing is maintained across all evaluated bandwidths, implying that the number of subcarriers, $P$, increases proportionally. The impact of signal bandwidth is more significant in TO estimation, as it enhances the resolution of delay estimation. The improvement in CFO estimation, shown in Fig. \ref{f5}(b), is attributed to the increased number of subcarriers, which provides additional processing gain over the delay axis, as indicated by (\ref{Eqn::49}).

\subsection{N-node Synchronization Performance of D-ISAC}
\begin{figure}[t!]
    \centering
    \subfloat[]{\includegraphics[scale=0.6]{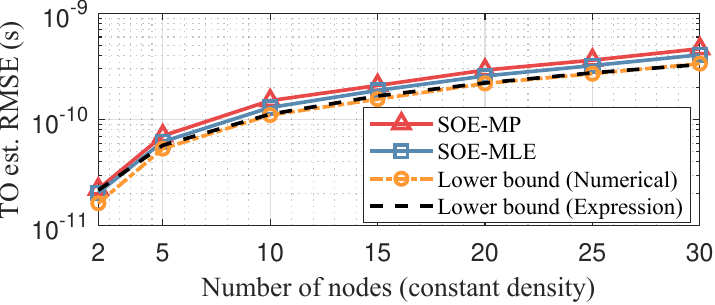}}  \\
    \centering
    \subfloat[]{\includegraphics[scale=0.6]{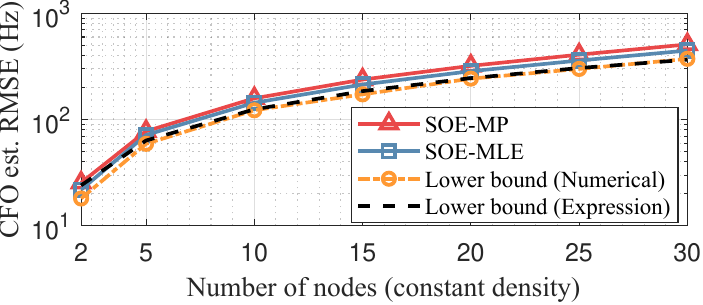}} 
    \caption{Total offset estimate errors in D-ISAC using the proposed SOE techniques with centered pairwise synchronization when the number of nodes increases while maintaining a fixed node density: (a) TO estimation, and (b) CFO estimation.}
    \label{f7}
\end{figure}

\begin{figure}[t!]
    \centering
    \subfloat[]{\includegraphics[scale=0.6]{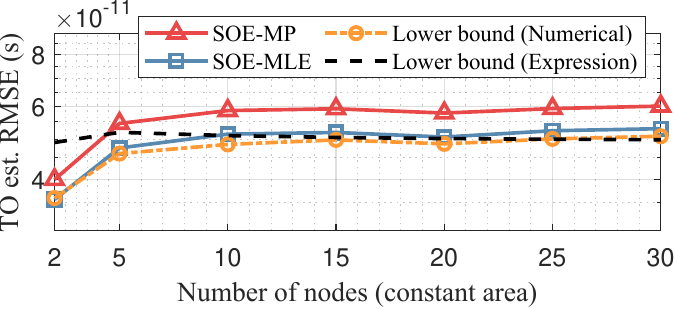}}  \\
    \centering
    \subfloat[]{\includegraphics[scale=0.6]{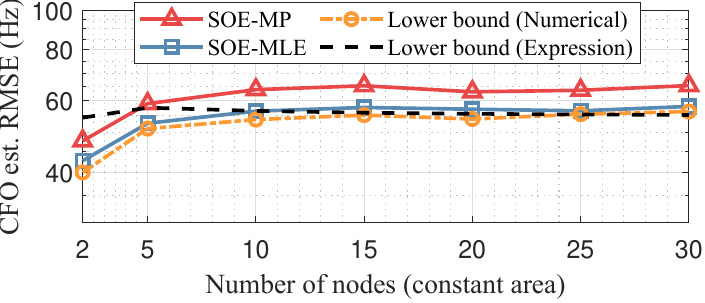}} 
    \caption{Total offset estimate errors in D-ISAC using the proposed SOE techniques with centered pairwise synchronization when the number of nodes increases while maintaining a fixed area size: (a) TO estimation, and (b) CFO estimation.}
    \label{f8}
\end{figure}
We validate the performance of the proposed synchronization technique over $N$ distributed nodes using CPS. In this simulation, a target is placed at the origin with an SNR of 17 dB at a 50 m distance in bistatic measurements. The ISAC nodes are randomly positioned within the area $A_E$, following the PPP with density $\mu = N/A_E$. First, by fixing the density at $\mu = 100$~$\text{nodes/km}^2$, the overall synchronization performance of D-ISAC is evaluated as the number of nodes increases, as shown in Fig. \ref{f7}. It is observed that both TO and CFO estimation errors increase with a larger number of nodes. This occurs because the SNR of the target, utilized for the proposed BSM, decreases as the average distance between the target and nodes increases. Additionally, the results verify that the total offset estimate variance expressions derived in Theorem \ref{theo2} closely match those obtained from Monte Carlo simulations.

Conversely, when the number of nodes increases within a fixed area of $(200)^2$ $\text{m}^2$, the total TO and CFO estimation errors remain constant, as shown in Fig. \ref{f8}. This is because the expected SNR of the target in bistatic measurements increases, leading to a reduction in offset estimation errors. Accordingly, the increase in offset estimate variance due to the larger number of nodes is compensated by the increase in node density, which is consistent with Theorem \ref{theo2}. Once again, the derived lower bound expressions align well with numerical simulation results as the number of nodes increases.

\subsection{D-ISAC Localization Performance Evaluation}
\begin{figure}[t!]
    \centering
    \subfloat[]{\includegraphics[scale=0.6]{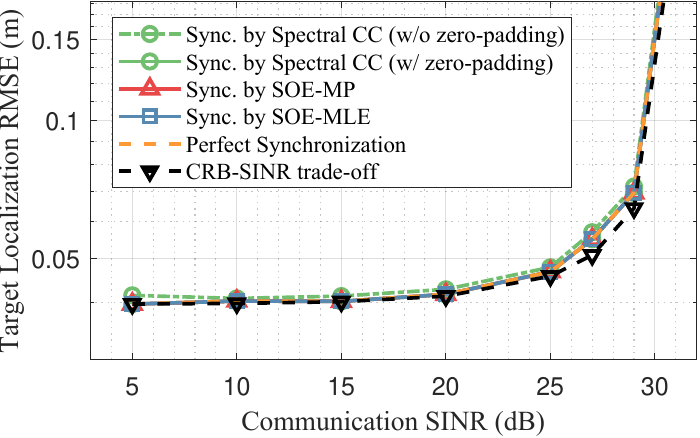}}  \\
    \centering
    \subfloat[]{\includegraphics[scale=0.6]{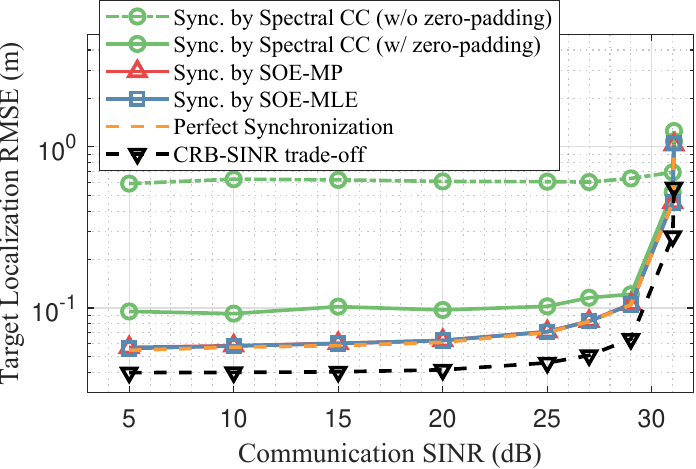}} 
    \caption{Target localization accuracy vs. communication SINR trade-offs in D-ISAC under different synchronization approaches: (a) D-ISAC performance with centralized processing, and (b) D-ISAC performance with decentralized processing for target localization. The proposed SOE approaches recover 96$\%$ of the bottom-line target localization performance with the fully-synchronous D-ISAC.}
    \label{f9}
\end{figure}

\begin{figure}[t!]
    \centering
    {\includegraphics[scale=0.6]{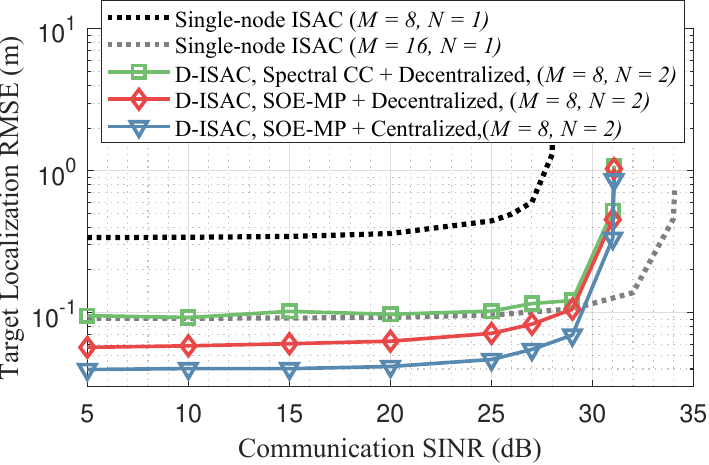}} 
    \caption{Performance gain of D-ISAC with various synchronization approaches compared to the single-node ISAC system. $M$ denotes the number of transmit antennas at each ISAC node.}
    \label{f10}
\end{figure}

\begin{figure}[t!]
    \centering
    {\includegraphics[scale=0.6]{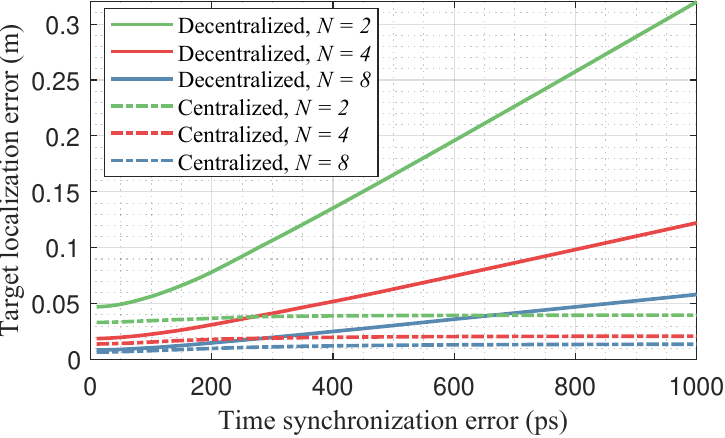}} 
    \caption{Target localization accuracy in D-ISAC with varying time-offsets between nodes.}
    \label{f11}
\end{figure}

In this section, we investigate the bottom-line D-ISAC performance under synchronization errors. Specifically, we consider a noncoherent D-ISAC system without phase-level synchronization, where target localization performance is significantly affected by time synchronization between nodes. Following \cite{han2025signaling}, we employ two ISAC nodes with a D-ISAC signaling design based on target localization CRB and communication SINR trade-offs, where each ISAC node is equipped with 8 transmit and receive antennas. The D-ISAC signaling is designed with the per-antenna constraint. A typical communication user, equipped with a single receive antenna, is positioned at \([0, -50 \, \mathrm{m}]\), while a single target is randomly placed within the ranges \([0 \, \mathrm{m}, 20 \, \mathrm{m}]\) and \([0 \, \mathrm{m}, 100 \, \mathrm{m}]\) with 25 dB SNR. For further details on the D-ISAC signaling design, please refer to \cite{han2025signaling}.

Fig. \ref{f9} presents the D-ISAC trade-off performance under different synchronization techniques. Centralized processing shows minimal degradation under synchronization errors, as monostatic measurements compensate for TO-induced errors in bistatic links, as seen in Fig. \ref{f9}(a). In contrast, decentralized processing, shown in Fig. \ref{f9}(b), is significantly affected due to reliance on a single monostatic link. D-ISAC with spectral CC exhibits a noticeable performance gap, while the proposed SOE-MLE and SOE-MP recover 96\% of the localization performance of a fully synchronized system, demonstrating their fine-grained offset estimation capability.

D-ISAC improves target localization by leveraging spatial diversity in distributed MIMO radar but is sensitive to synchronization errors. Fig. \ref{f10} compares noncoherent D-ISAC with a single-node ISAC system. When using the same number of antennas per node, D-ISAC achieves better localization accuracy and SINR due to power combining. However, a single-node ISAC system with an equivalent total number of antennas attains higher SINR due to coherent gain. While D-ISAC benefits from spatial diversity, spectral CC-based synchronization (250 ps accuracy) prevents it from outperforming the single-node system. In contrast, the proposed SOE-MP technique (10 ps accuracy) provide enough cooperative sensing gain even with the decentralized processing.

To quantify target localization error under time offsets in D-ISAC, we evaluate the target localization HCRB using (\ref{Eqn::64}). Fig. \ref{f11} illustrates the impact of synchronicity on D-ISAC performance for centralized and decentralized processing, assuming node positions follow a PPP with $\mu = 100$~$\text{nodes/km}^2$. Centralized processing remains robust against synchronization errors, whereas decentralized processing suffers increasing localization error with higher TO, though the effect diminishes as node density increases. More nodes provide independent bistatic measurements, enhancing spatial diversity and mitigating the impact of synchronization errors through signal fusion.

\section{Conclusion}
This paper proposed the over-the-air time-frequency synchronization framework for D-ISAC, leveraging reciprocal bistatic sensing channels for fine-grained TO and CFO estimation. By introducing the OR property and BSM procedure, the proposed SOE techniques significantly outperformed conventional spectral CC-based methods, achieving synchronization accuracy close to theoretical lower bounds. Furthermore, the framework was extended to multi-node D-ISAC systems using CPS, providing a theoretical bound on synchronization performance with randomly deployed nodes. Numerical simulations validated the effectiveness of the proposed techniques, confirming their fine-grained offset estimation performance. Additionally, the results revealed the impact of synchronization accuracy on overall D-ISAC performance, demonstrating its advantages in improving target localization in D-ISAC systems, which recovers up to 96$\%$ of the bottom-line target localization performance of the fully-synchronous D-ISAC. The proposed framework offers an efficient solution for synchronization in D-ISAC systems, supporting enhanced sensing and communication capabilities in future wireless networks. Future work includes exploring phase-level synchronization for coherent D-ISAC systems, which can further enhance D-ISAC performance.

\appendices
\section{Proof of Lemma \ref{lemma1}}
\label{proof_lemma}
Without loss of generality, let us assume that $R_{1} \leq R_{2} \leq \cdots \leq R_{N}$ and consider the $n^{th}$ node as the reference node. Substituting (\ref{Eqn::48}) and (\ref{Eqn::52}) into (\ref{Eqn::50}), the lower bound of the total offset estimate variance for a sufficiently large target SNR simplifies to
\begin{align}
    \widetilde{\sigma}_{\Delta \delta_{t}}^2 & \geq \mathbb{E} \left[\sum_{i = 1, i \neq n}^{N}\frac{3 \sigma^2 R_{n}^{2} R_{i}^{2}}{4 \pi^2 \eta \alpha (\Delta f)^2  P^3 Q}\right] \nonumber \\
    & = \frac{3 \sigma^2}{4 \pi^2 \eta \alpha (\Delta f)^2  P^3 Q} \mathbb{E} \left[ \sum_{i = 1, i \neq n}^{N}  R_{n}^{2} R_{i}^{2}\right]. \label{Eqn::53}
\end{align}
Since $R_{1}, R_{2}, \dots, R_{N}$ are independent variables, it follows that
\begin{align}
\mathbb{E} \left[ \sum_{i = 1, i \neq n}^{N}  R_{n}^{2} R_{i}^{2}\right] = \mathbb{E} \left[R_{n}^{2}\right] \mathbb{E} \left[\sum_{i = 1, i \neq n}^{N} R_{i}^{2}\right]. \label{Eqn::53-2}
\end{align}
The expectation $\mathbb{E} \left[R_{i}^{2}\right]$ can be computed as
\begin{align}
\mathbb{E} \left[ R_i^2 \right] & = \int_0^\infty r^2 f_{R_i}(r) \, dr \nonumber \\
& = \frac{i}{\mu \pi}, \quad \forall i = 1, 2, \dots, N, \label{Eqn::54}
\end{align}
where the probability density function (PDF) of $R_i$ is given by $f_{R_i}(r) = \frac{2 (\mu \pi r^2)^{(i-1)}}{(i-1)!} \mu \pi r e^{-\mu \pi r^2}, \quad r \geq 0.$
Accordingly, the term $\mathbb{E} \left[R_{n}^{2}\right] \mathbb{E} \left[\sum_{i = 1, i \neq n}^{N} R_{i}^{2}\right]$ is minimized when $n = 1$. Thus, the closest node to the target is the optimal reference node, completing the proof.

\section{Proof of Theorem \ref{theo2}}
\label{proof2}
With the reference node selected as $n = 1$, (\ref{Eqn::53-2}) simplifies to $\mathbb{E} \left[ \sum_{n = 2}^{N}  R_{1}^{2} R_{n}^{2}\right]$. Using (\ref{Eqn::54}), this expectation can be further expressed as
\begin{align}
    \mathbb{E} \left[ \sum_{n = 2}^{N}  R_{1}^{2} R_{n}^{2}\right] = \frac{(N-1)(N+2)}{2(\mu \pi)^2}. \label{Eqn::57}
\end{align}
Substituting (\ref{Eqn::57}) into (\ref{Eqn::53}), we obtain the final lower bound for TO estimation variance as
\begin{align}
    \widetilde{\sigma}_{\Delta \delta_{t}}^2 & \geq \frac{3 \sigma^2 (N-1) (N+2)}{8 \pi^4 \eta \alpha (\Delta f)^2  \mu^2 P^3 Q}.
\end{align}
The proof for (\ref{Eqn::56}) follows the same approach as for TO estimation.
\ifCLASSOPTIONcaptionsoff
  \newpage
\fi



\bibliographystyle{IEEEtran}
\bibliography{IEEEabrv,reference}
%



%





\vfill


\end{document}